\documentclass[a4paper,11pt]{article}
\pdfoutput=1
\usepackage{jheppub}
\usepackage[T1]{fontenc}

\usepackage{physics}
\usepackage{subcaption}
\usepackage{afterpage}

\newcommand\del{\partial}

\newcommand\Abs{\text{Abs}}
\newcommand{\taus}{{\tau_1\tau_2\tau_3\tau_4}}

\def\L{\mathcal{L}}
\def\K{\mathcal{K}}
\def\T{\mathcal{T}}
\def\H{\mathcal{H}}
\def\M{\mathcal{M}}

\def\d{\mathrm{d}}
\newcommand\ptwiddle[1]{\mathord{\mathop{#1}\limits^{\scriptscriptstyle(\sim)}}}

\title{\boldmath Positivity bounds in vector theories}

\author[a,b,c]{Claudia de Rham,}
\author[d]{Laura Engelbrecht,}
\author[c,d,e]{Lavinia Heisenberg,}
\author[d]{Alice L\"uscher}

\affiliation[a]{Theoretical Physics, Blackett Laboratory, Imperial College, London, SW7 2AZ, UK}
\affiliation[b]{CERCA, Department of Physics, Case Western Reserve University, 10900 Euclid Ave, Cleveland,
OH 44106, USA}
\affiliation[c]{Perimeter Institute for Theoretical Physics,
31 Caroline St N, Waterloo, Ontario, N2L 6B9, Canada}
\affiliation[d]{Institute for Theoretical Physics,
ETH Z\"urich, Wolfgang-Pauli-Strasse 27, 8093, Z\"urich, Switzerland}
\affiliation[e]{Institute for Theoretical Physics, Heidelberg University, Philosophenweg 16, 69120 Heidelberg, Germany}

\emailAdd{c.de-rham@imperial.ac.uk}
\emailAdd{ljohnson@phys.ethz.ch}
\emailAdd{laviniah@ethz.ch}
\emailAdd{aluescher@ethz.ch}

\abstract{Assuming unitarity, locality, causality, and Lorentz invariance of the, otherwise unknown, UV completion, we derive a new set of constraints on the effective field theory coefficients for the most general, ghost-free Generalized Proca and Proca Nuevo massive vector models.
For the Generalized Proca model, we include new interactions that had not been previously considered in the context of positivity bounds and find these additional terms lead to a widened parameter space for the previously considered interactions. Although, the Generalized Proca and Proca Nuevo models are inequivalent, we find interesting analogues between the coefficients parameterizing the two models and the roles they play in the positivity bounds.}

\begin{document}

\hspace{4.2in} \mbox{Imperial/TP/2022/CdR/02}\\\vspace{1.53cm} 

\maketitle
\flushbottom

\section{Introduction}
Effective field theories (EFTs), describe low energy (IR) physics, up to a certain cutoff $\Lambda$ beyond which additional ingredients needs to be considered. Upon constructing an EFT, every operator consistent with the field content and the symmetry of the system should in principle be included with an arbitrary Wilsonian coefficient, leading to parameter spaces which may be typically be significant.  Consider for instance the Standard Model EFT (SMEFT), just accounting for up to dimension-6 baryon-number conserving operators already leads to a 59-dimensional phase space \cite{Buchmuller1986621,Grzadkowski:2010es}.
When considering experiments or observations where searching large parameter spaces may be costly or infeasible, narrowing down the parameter space can be invaluable, allowing for more targets searches or even allowing one to theoretically rule out certain theories as incompatible with a standard UV completion. Even if the full (UV complete) model at high energies is unknown, the parameter space of an EFT may be restricted using constraints arising from hypotheses on the UV completion.

The physical requirements of unitarity, locality, causality, and Lorentz invariance of the UV completion of an EFT translate into unitarity, analyticity and crossing symmetry of scattering amplitudes. Put together, these properties give rise to a dispersion relation, whose positivity constrains the amplitudes.
Schematically, analyticity allows one to write the amplitude $\mathcal{A}$ in the complex plane of the Mandelstam center of mass energy square variable $s$ as a contour integral and to deform the contour to obtain an integral of the imaginary part, which is positive by unitarity (optical theorem) and crossing relations,
\begin{equation}
    \frac{1}{2}\mathcal{A}''(s) = \frac{1}{2\pi i}\oint_\mathcal{C} \d s'\frac{\mathcal{A}(s')}{(s'-s)^3} =\frac{1}{\pi} \int_\text{cuts}\d\mu\frac{\Im A(\mu)}{(\mu-s)^3}>0\, .
\end{equation}
In practice, $\mathcal{A}''(s)$ is computed in a given EFT, whereas the last inequality is not usually computed explicitly, but its positivity is guaranteed by the assumptions on the UV completion and in turn ensures the positivity of the $\mathcal{A}''(s)$. This procedure provides positivity bounds which can be used to constrain the parameter space of EFTs.

It has long been known that analyticity and dispersion relations lead to positivity constraints, but it is only quite recently that \cite{Adams_2006} exploited these constraints to bound EFT coefficients. Since their foundational work, which applied to scalar theories in the forward limit, some extensions have been worked out both away from the forward limit \cite{de_Rham_2017} and for spinning particles \cite{Bellazzini2017,de_Rham_2018,davighi2021natural}. These bounds have been applied to massive particles of spin-1 \cite{Bonifacio_2016,de_Rham_2017gal,de_Rham_2019}, as well as on massive spin-2 fields \cite{Cheung2016,Bonifacio_2016,Bellazzini:2017fep,de_Rham_2018improved,de_Rham_2019,Alberte2020,Alberte20202,Wang2021}.
In particular it was shown in \cite{Cheung2016,de_Rham_2019,Alberte2020,Alberte20202} how involving bounds with mixed field polarization can significantly reduce the allowed region of parameter space and lead to compact bounds.
Recently, the parameter space has been carved in an even more systematical way making use of clever non-linear bounds \cite{Bellazzini_2021,Arkani-Hamed:2020blm,Chiang2021}.
Insight from full crossing symmetric dispersion relations as highlighted in \cite{Sinha:2020win,Haldar:2021rri,Raman:2021pkf,Chowdhury:2021ynh,Sinha:2022sdo} was for instance folded into new types of compact bounds in \cite{tolley2021new,Caron-Huot2021} with applications to multi-field EFTs in \cite{Du_2021} and EFTs involving higher spin fields in \cite{Chowdhury:2021ynh}.
Using these methods, the massless and massive Galileon have been found to admit no standard UV completion \cite{Adams_2006,tolley2021new}. Likewise, in \cite{Bonifacio_2016}, it was shown that the simplest vector Galileon model also cannot admit a standard UV completion. Nonetheless, as argued in \cite{Cheung2016,de_Rham_2018improved,de_Rham_2019}, such conclusions do not prevent massive gravity (or GP, PN) to possibly admit one, even though the massive Galileon emerges as their helicity-0 mode in the decoupling limit. Moreover, a violation of the bounds may  not be dramatic in itself. It simply implies that if a UV completion of the model is to exist, it may not be enjoy the same level of locality as required in the derivation of positivity bounds (and in particular a violation of the Froissart-like bound may occur) as for instance illustrated in \cite{Keltner:2015xda} or may occur in other UV finite and unitary models \cite{Tolley:2019nmm}.

Due to their relevance in modern physics, some EFTs have been particularly popular for the application of the bounds. Positivity bounds have been extensively studied in the case of (massless) gravity, where both pure gravity \cite{Bern_2021,Chiang2022,CaronHuot2022,Herrero_Valea_2021} and other fields in the gravitational context \cite{Alberte_2020spin2,Tokuda_2020,Alberte2021,Alberte_202222,Caron_Huot_2021GR}. Positivity bounds for gravity require special care due to its massless nature and the presence of a pole in the $t$-channel exchange, leading to additional complications. Moreover, positivity bounds in gravity can be related to the Weak Gravity Conjecture \cite{Cheung_2014,Hamada_2019,Arkani_Hamed_2022,Henriksson2022}. The Standard Model EFT  has also drawn a great deal of attention in the positivity bounds community  \cite{Zhang_2019,Zhou2019,Bellazzini_2018SM,Remmen_2019,Zhang_2020,Zhou2021,Fuks_2021,Remmen_2020,Bonnefoy_2021,Chala2021}. Constraining the SMEFT parameter space makes both the experimental searches and the theoretical interpretations of the data more efficient. If some violations of the bounds were detected, it would indicate that some assumptions on the UV-complete theory are erroneous. The application of the bounds in cosmology, as conducted in \cite{Melville_2020,Noller2021,Kim_2019,Herrero_Valea_2019,Ye_2020,melville2022positivity,Creminelli:2022onn}, can be combined with observational data to strongly constrain the parameter space of dark energy and modified gravity models.
(See \cite{deRham:2022hpx} for a recent review of the positivity bounds and their applications.)

The aim of this work is to provide new constraints on vector EFTs via the positivity bounds technology.
Positivity bounds on certain Proca interactions have been considered in \cite{Bonifacio_2016, de_Rham_2019}, but these were not the most general interactions allowed within these models. We consider the most general possible parity even interaction terms allowed for Generalized Proca theories which include terms with scalar-vector mixing in the decoupling limit. Positivity bounds on Proca Nuevo models have never previously been considered and are studied here for the first time.

We begin by presenting explicitly the two massive vector models of interest, the Generalized Proca \cite{Heisenberg2014} and Proca-Nuevo \cite{derham2020} in section~\ref{sec.models}. Then, we review the spinning positivity bounds from \cite{de_Rham_2018} in section~\ref{sec.bounds}. The reader familiar with these topics may skip these first two sections. Finally, we present our results for the constraints on the two Proca theories in section~\ref{sec.results}.
We work in a $3+1$ dimensional flat space using the mostly-plus signature $(-,+,+,+)$.

\section{Vector Models}\label{sec.models}
 There were initial attempts to write down derivative self-interactions for a massless Abelian vector field, but it was concluded that the Maxwell term is the only possible term compatible with gauge invariance and leading to second order equation of motion (EOM) on flat space, hence a no go theorem for massless vector Galileons \cite{nogo2014}\footnote{Even though a non-minimal coupling of the gauge-invariant field strength to gravity exists, for most of the relevant applications it is unstable \cite{BeltranJimenez:2013btb}.}. On the other hand, it is possible to construct derivative self-interactions with second order EOM if we give up gauge symmetry and consider massive vector theories, also called Generalized Proca theories. The Generalized Proca (GP) model \cite{Heisenberg2014,Tasinato2013,Allys2016,Jimenez2016} is the most general Abelian self-interacting massive spin-1 theory\footnote{A non-Abelian version of the Generalized Proca has been investigated in \cite{BeltranJimenez:2016afo,Allys_2016}.} with second order EOM, even if the interaction terms from the Lagrangian appear to have higher order derivatives (see \cite{Heisenberg:2018vsk} for a review). Moreover, it enjoys a Vainshtein screening mechanism, making it compatible with solar system tests of gravity \cite{De_Felice_2016}.
These models have found various phenomenological
applications in astrophysics \cite{Chagoya_2017,Kase_2018,Kase_2020,Garcia_Saenz_2021,Brihaye_2022} notably for black holes  \cite{Chagoya_2016,Minamitsuji_2016,Cisterna_2016,Chagoya_2017,Heisenberg_2017,Minamitsuji2017,Heisenberg_20172,Kase_20182,Kase_20183,Rahman_2019}, and in cosmology  \cite{Felice_2016,De_Felice_20162,Heisenberg_2016,Nakamura_2017,Emami_2017,De_Felice_2017,Kase_20184,Dom_nech_2018,Nakamura_2019,Oliveros_2019,Felice_2020,Heisenberg_2021}.

In order to avoid instabilities, a massive vector theory should propagate three degrees of freedom. This is achieved by imposing a constraint, such as the degeneracy of the Hessian matrix. In the Generalized Proca model, this constraint manifests itself by the non-propagation of the temporal component, and the absence of Ostrogradski instability is made apparent by the second order EOM. However, there is no reason that the constraint may not be realized in another way (in fact see \cite{Heisenberg:2016eld} and \cite{BeltranJimenez:2019wrd}).

Another kind of self-interacting massive vector theory has recently been discovered \cite{derham2020}. In fact, a Proca theory different from GP emerges from the decoupling limit of massive gravity on AdS space \cite{Laura2018}, and inspired the construction of a new Proca theory on flat space, dubbed Proca-Nuevo (PN) \cite{derham2020}. As intended, this model propagates three degrees of freedom, but realizes the constraint non-linearly as opposed to GP, where the constraint is realized order by order.
Moreover, both the GP and PN theories lead to a time-dependent vector condensate which could play the role of a dark energy fluid, driving the accelerated expansion of the universe that we currently observe. Additionally, these models have a technically natural vector mass and dark energy scale \cite{Zosso2021,derham2021quantum}. The
application of an extended PN model in cosmology have been successfully considered in \cite{derham2021cosmology,Pozsgay2022}.
\subsection{Generalized Proca}\label{sec.GP}

The Generalized Proca interactions \cite{Heisenberg2014}, sometimes referred to as vector Galileons, are massive vector self-interactions built out of two requirements. One, that the equations of motion are second order. Two, that the temporal component is not dynamical. In turn, these guarantee that only three healthy degrees of freedom propagate.

A way to construct this theory is to write down all possible interactions at each order in derivatives and tune the coefficients to have a degenerate Hessian \cite{Heisenberg2014}. This procedure produces the following model for the massive vector field $A_\mu$
\begin{equation}
	\L_{\text{GP}} = \sum_{n=2}^6 \L_n\,,
\label{eq:LGP}
\end{equation}
with
\begin{equation}\label{eq.lgp}
    \begin{aligned}
	\L_2 &= f_2(A_{\mu}, F_{\mu\nu}, \tilde{F}_{\mu\nu})\\
	\L_3 &= f_3(A^2) (\partial \cdot A)\\
	\L_4 &= f_4(A^2) [(\partial \cdot A)^2 - \partial_{\mu} A_{\nu} \partial^{\nu} A^{\mu}]\\
	\L_5 &= f_5(A^2) [(\partial \cdot A)^3 -3 (\partial \cdot A) \partial_{\mu} A_{\nu} \partial^{\nu} A^{\mu} + 2 \partial_{\mu} A_{\nu} \partial^{\nu} A^{\rho} \partial_{\rho} A^{\mu} ] \\ & + \tilde{f}_5(A^2) \tilde{F}^{\mu \alpha} \tilde{F}^{\nu}_{\phantom{\nu} \alpha} \partial_{\mu} A_{\nu}\\
	\L_ 6 &= \tilde{f}_6(A^2) \tilde{F}^{\mu \nu} \tilde{F}^{\alpha \beta} \partial_{\alpha} A_{\mu} \partial_{\beta} A_{\nu}\, ,
\end{aligned}
\end{equation}
where $F_{\mu\nu}=\del_\mu A_\nu-\del_\nu A_\mu$, and $\Tilde{F}^{\alpha\beta}=\frac{1}{2}\epsilon^{\alpha\beta\mu\nu}F_{\mu\nu}$ is the dual field-tensor. The $f_n$'s are arbitrary functions of their arguments. In particular, $f_2$ contains the kinematic and mass terms. This theory is the most general one respecting the two requirements. Any other possible interactions are related to these by total derivatives and disformal transformations of the metric.

Moreover, expanding the $f_n(A^2)$ in power series of $A^2$, the full Lagrangian can be expressed perturbatively as
\begin{equation}
    \mathcal{L}_{\text{GP}}=\sum_{n=2}^\infty\frac{1}{\Lambda_2^{2(n-2)}}\mathcal{L}_{\text{GP}}^{(n)}\, ,
\end{equation}
where each $\mathcal{L}^{(n)}$ contains the $n$-point interactions, and $\Lambda_p=(m^{p-1}M_{\text{Pl}})^{1/p}$ is a dimensionful scale. In this work, we are interested in computing tree-level 2-2 scattering amplitudes, such that it is sufficient to consider only up to quartic interactions. The first terms of the perturbative expansion can be expressed as
\begin{equation}\label{eq:LGPpert}
\begin{aligned}
\mathcal{L}_{\text{GP}}^{(2)}&=-\frac{1}{4}F^{\mu\nu}F_{\mu\nu}-\frac{1}{2}m^2 A^2 \\
    \mathcal{L}_{\text{GP}}^{(3)}&=a_1m^2A^2\del_\mu A^\mu+ a_2\Tilde{F}^{\mu\alpha}\Tilde{F}^\nu_{\  \alpha}\del_\mu A_\nu \\
     \mathcal{L}_{\text{GP}}^{(4)}&=b_1 m^4A^4+b_2m^2A^2F_{\mu\nu}^2+ b_3m^2A^2[(\del\cdot A)^2-\del_\mu A_\nu\del^\nu A^\mu]+b_4m^2A_\mu A^\nu F^{\alpha\mu}F_{\alpha\nu} \\ &\, + b_5F^{\mu\nu}F^{\alpha\beta}F_{\mu\alpha}F_{\nu\beta}+b_6(F_{\mu\nu}^2)^2
    +b_7 \Tilde{F}^{\alpha\beta}\Tilde{F}^{\mu\nu}\del_\alpha A_\mu\del_\beta A_\nu \, ,
\end{aligned}
\end{equation}
where the coefficients $a_i$ and $b_i$ are dimensionless coupling constants. There exist different ways to define the theory perturbatively, but they only differ by field redefinitions and total derivatives, and are therefore equivalent. In any formulation, there are respectively 2 and 7 independent terms at cubic and quartic orders. In this work, we follow the formulation of \cite{derham2020}, which offers a good comparison with the Proca-Nuevo model.

In order to study the decoupling limit of the theory, we perform the Stückelberg procedure, introducing the Stückelberg field $\phi$,
\begin{equation}
    A_\mu\to A_\mu+\frac{1}{m}\del_\mu \phi\, ,
\end{equation}
such that, in the decoupling limit, the 3 helicities ($\lambda=\pm1,0$) of the massive vector decomposes into 2 of a massless vector $A_\mu$ (helicity-1 modes) and 1 of a massless scalar $\phi$ (helicity-0 mode). Then, taking the decoupling limit corresponds to send the mass to zero and the scale to infinity while keeping the lowest interaction scale constant:
\begin{equation}\label{eq.DL}
    m\to0, \quad \Lambda_2\to\infty,\quad \text{while} \quad \Lambda_3\equiv(m\Lambda_2^2)^{1/3}=\text{const}.
\end{equation}
In these limits, the perturbative Lagrangian reads
\begin{equation}
    \mathcal{L}_{\text{DL GP}}=\sum_{n=2}^\infty\frac{1}{\Lambda_3^{3(n-2)}}\mathcal{L}_{\text{DL GP}}^{(n)}\, ,
\end{equation}
with
\begin{equation}\label{eq.GPDL}
\begin{aligned}
\mathcal{L}_{\text{DL GP}}^{(2)}&=-\frac{1}{4}F^{\mu\nu}F_{\mu\nu}-\frac{1}{2}(\del\phi)^2 \\
    \mathcal{L}_{\text{DL GP}}^{(3)}&=a_1(\del\phi)^2\Box\phi+ a_2\Tilde{F}^{\mu\alpha}\Tilde{F}^\nu_{\  \alpha}(\del_\mu\del_\nu\phi) \\
     \mathcal{L}_{\text{DL GP}}^{(4)}&=b_3(\del\phi)^2[(\Box\phi)^2-(\del_\mu\del_\nu\phi)^2]
    +b_7 \Tilde{F}^{\alpha\beta}\Tilde{F}^{\mu\nu}(\del_\alpha \del_\mu\phi)(\del_\beta\del_\nu\phi).
\end{aligned}
\end{equation}
$\L^{(2)}$ contains the decoupled kinetic terms of the massless vector and scalar modes. The terms in $a_1$ and $b_3$ correspond respectively to the cubic and quartic Galileon interactions, whereas those in $a_2$ and $b_7$ mix the scalar and vector modes.
In particular, \cite{de_Rham_2019,Bonifacio_2016} use models that only gives Galileons in the decoupling limit, i.e.\ with $a_2$, $b_7=0$ to derive some positivity bounds for the Generalized Proca.

\subsection{Proca-Nuevo}\label{sec.PN}
The Proca-Nuevo model \cite{derham2020} is built from the same determinant formulation as dRGT massive gravity \cite{deRham:2010kj}. There, the reference metric is defined as $f_{\mu\nu}=\del_\mu\phi^a\del_\nu\phi^b\eta_{ab}$, where $a$ is a Lorentz index, such that the quadruplet of Stückelberg fields $\phi^a$ is identified as a Lorentz vector, which can subsequently be decomposed to
\begin{equation}
    \phi^\mu=x^\mu+\frac{1}{\Lambda_2^2}A^\mu\, ,
\end{equation} for a vector field $A_\mu$. Then, in terms of this vector field, $f_{\mu\nu}$ reads
\begin{equation}\label{eq.fofA}
    	f_{\mu\nu}[A] = \eta_{\mu\nu} + 2 \frac{\del_{(\mu} A_{\nu)}}{\Lambda_2^2} + \frac{\del_{\mu} A_\alpha \del_{\nu} A_\beta \eta^{\alpha \beta}}{\Lambda_2^4}\,.
\end{equation}
Next, we keep the same deformed determinant as in massive gravity,
but with $g_{\mu\nu}=\eta_{\mu\nu}$ the flat Minkowski metric and $f_{\mu\nu}$ expressed in terms of the vector field as defined in \eqref{eq.fofA},
\begin{equation}
    \K^\mu_{\ \nu}=\left(\sqrt{\eta^{-1}f[A]}\right)^\mu_{\ \nu}-\delta^\mu_{\ \nu}\, .
\end{equation}
Finally, the full Lagrangian of this theory, defined by $\det(\delta^\mu_{\ \nu}+\K^\mu_{\ \nu})$, is given by
\begin{equation}
	\L_{\text{PN}} = \Lambda_2^4\sum_{n=0}^4 \alpha_n(A^2)\L_n[\K]\,,
\end{equation}
with $\L_n[\K]$, the elementary Lagrangians
\begin{equation}
    \begin{aligned}
	\L_0 &= 4!\\
	\L_1 &= 3![\K] \\
	\L_2 &= 2!([\K]^2 - [\K^2]) \\
	\L_3 &= [\K]^3 - 3[\K][\K^2] + 2[\K^3] \\
	\L_4 &= [\K]^4 - 6[\K]^2[\K^2] + 3[\K^2]^2 + 8[\K][\K^3] - 6[\K^4] \,.
\end{aligned}
\end{equation}
Note that, despite its similar construction to massive gravity, by the definition \eqref{eq.fofA} of $f_{\mu\nu}$ in terms of a vector field, this model contains no tensor degrees of freedom. It is a pure vector model, with an infinite tower of self-interactions. These interactions lead to higher order equation of motion, such that one may worry about the potential propagation of an Ostrogradski ghost. However, the authors of \cite{derham2020} exhibited a null eigenvector of the Hessian, implying the presence of a constraint that removes the ghostly degree of freedom. Therefore, the above construction gives rise to a self-interacting massive vector theory which propagates only three healthy degrees of freedom. This theory is called Proca-Nuevo.

Additionally, the Proca-Nuevo theory is inequivalent to the Generalized Proca. This can be proved explicitly by trying to match the scattering amplitudes of both theories, as was done in \cite{derham2020}. This is not in contradiction with the uniqueness of GP as the defining hypotheses differ. Indeed, GP is characterized by having second order equations of motion, while PN violates this condition. Moreover, the realisation of the constraint is imposed at each order in GP, whereas all orders are needed for PN to realise it.
As a consequence of this last point, the null eigenvectors of the two theories are fundamentally different, such that the constraint can not be simultaneously realised \cite{derham2020}.
Therefore, a model including both the GP and PN interactions would suffer from ghost instabilities unless some non-trivial restrictions are made (at least in the absence of gravity) \cite{derham2021cosmology}.

Furthermore, the PN Lagrangian can also be written perturbatively,
\begin{equation}\label{eq.LPNpert}
    \mathcal{L}_{\text{PN}}=\sum_{n=2}^\infty\frac{1}{\Lambda_2^{2(n-2)}}\mathcal{L}_{\text{PN}}^{(n)}\,,
\end{equation}
where the expressions for $\L^{(n)}$ are obtained by expanding the arbitrary functions $\alpha_n$ in power of $A^2$
\begin{equation}
    \alpha_n(A^2)=\bar\alpha_n+\frac{m^2}{\Lambda_2^4}\bar\gamma_n A^2+\frac{m^4}{\Lambda_2^6}\bar\lambda_n A^4+\dots
\end{equation}
The dimensions are contained in the scale $\Lambda_2$, while the coefficients are dimensionless.
In order to get the usual normalization for the Maxwell and mass terms, we set $\bar\alpha_1=-\frac{1}{3}(1-2\bar\alpha_2)$ and $ \bar\gamma_0=-\frac{1}{48}$.
We observe that the combination $(1+4\bar\alpha_2-6\bar\alpha_3)$ appears repeatedly in the results and thus redefine $\bar\alpha_2$ accordingly. We also rescale the coefficients $\bar\gamma_1$, $\bar\gamma_2$, and $\bar\lambda_0$ so that they later compare nicely to the GP parameters $a_1$, $b_3$, and $b_1$ respectively. In sum, we define the following parameters
\begin{equation}\label{eq.PNredef}
\begin{aligned}
       &\alpha_2'=1+4\bar\alpha_2-6\bar\alpha_3, \quad \alpha_3'=3(\bar\alpha_3-4\bar\alpha_4), \\ &\gamma_1'=6\bar\gamma_1, \quad \gamma_2'=2\bar\gamma_2, \quad \lambda_0'=24\bar\lambda_0.
\end{aligned}
\end{equation}
Under this redefinition, the perturbative PN Lagrangian \eqref{eq.LPNpert} reads
\begin{equation}\label{eq.LPNredef1}
\begin{aligned}
&\mathcal{L}_{\text{PN}}^{(2)}=-\frac{1}{4}F^{\mu\nu}F_{\mu\nu}-\frac{1}{2}m^2 A^2 \\
    &\mathcal{L}_\text{PN}^{(3)}= \gamma_1' m^2 A^2\del_\mu A^\mu+\frac{1}{8}(\alpha_2'-1)[F^2][\del A]+\frac{1}{4}(2-\alpha_2')F^2_{\mu\nu}\del^\mu A^\nu \\
    &\mathcal{L}_\text{PN}^{(4)}= \lambda_0' m^4 A^4 +m^2 A^2\Big[\gamma_2'[\del A]^2-\frac{1}{2}\left(\frac{1}{2}\gamma_1'+\gamma_2'\right)\del_\mu A_\nu \del^\nu A^\mu +\frac{1}{2}\left(\frac{1}{2}\gamma_1'-\gamma_2'\right)\del_\mu A_\nu \del^\mu A^\nu \Big] \\ & \qquad + \frac{1}{128}(-1+\alpha_2'-2\alpha_3')[F^2]^2+\frac{1}{64}(10-5\alpha_2'-14\alpha_3')F^2_{\mu\nu}F^{2\mu\nu} \\ & \qquad+\frac{1}{8}\Big[\alpha_3'[F^2]([\del A]^2-\del_\mu A_\nu\del^\nu A^\mu)+\left(-2+\alpha_2'+4\alpha_3'\right)F^{2\mu\nu}\del^\alpha A_\mu\del_\alpha A_\nu \\& \qquad\qquad +\left(1-\alpha_2'-4\alpha_3'\right)F^{2\mu\nu}[\del A]\del_\mu A_\nu +\left(-2+\alpha_2'+2\alpha_3'\right)F^{\mu\nu}F^{\alpha\beta}\del_\mu A_\alpha\del_\nu A_\beta\Big],
\end{aligned}
\end{equation}
where we used the notation $F^2_{\mu\nu}=F_\mu^{\ \alpha}F_{\nu\alpha}$ and $[F^2]=F^{\mu\nu}F_{\mu\nu}$.
Following this reformulation, all of the combinations of the parameters $\{\bar\alpha_2,\bar\alpha_3,\bar\alpha_4\}$ in the Lagrangian reduce to combinations of only $\{\alpha_2',\alpha_3'\}$.
We mentioned before that the PN interactions have to be related in a specific way for the constraint to be respected. This is well illustrated in the perturbative Lagrangian, where the coefficients do not correspond to a specific interaction, but rather intertwine them. This is in contrast with GP in \eqref{eq.GPDL} where each interaction has its own coefficient.

Finally, taking the decoupling limit \eqref{eq.DL} of this theory gives
\begin{equation}
    \mathcal{L}_{\text{DL PN}}=\sum_{n=2}^\infty\frac{1}{\Lambda_3^{3(n-2)}}\mathcal{L}_{\text{DL PN}}^{(n)}\, ,
\end{equation}
with
\begin{equation}\label{eq.PNDL}
\begin{aligned}
&\mathcal{L}_{\text{DL PN}}^{(2)}=-\frac{1}{4}F^{\mu\nu}F_{\mu\nu}-\frac{1}{2}(\del\phi)^2 \\
    &\mathcal{L}_{\text{DL PN}}^{(3)}=\gamma_1'(\del\phi)^2\Box\phi+\frac{1}{8}(\alpha_2'-1)[F^2]\Box\phi+\frac{1}{4}(2-\alpha_2')F^2_{\mu\nu}(\del^\mu\del^\nu\phi)\\
     &\mathcal{L}_{\text{DL PN}}^{(4)}=\gamma_2'(\del\phi)^2[(\Box\phi)^2-(\del_\mu\del_\nu\phi)^2] \\ & \:+\frac{1}{8}\Big[\alpha_3'[F^2][(\Box\phi)^2-(\del_\mu\del_\nu\phi)^2]+\left(-2+\alpha_2'+4\alpha_3'\right)F^{2\mu\nu}(\del^\alpha \del_\mu\phi)(\del_\alpha \del_\nu\phi) \\ & \: +\left(1-\alpha_2'-4\alpha_3'\right)F^{2\mu\nu}\Box\phi(\del_\mu\del_\nu\phi)+\left(-2+\alpha_2'+2\alpha_3'\right)F^{\mu\nu}F^{\alpha\beta}(\del_\mu\del_\alpha\phi)(\del_\nu\del_\beta\phi)\Big]\, .
\end{aligned}
\end{equation}
The scalar modes are the $\bar\gamma_1$ and $\bar\gamma_2$ terms and correspond to the cubic and quartic Galileon interactions respectively. The other terms match the vector-scalar sector of the decoupling limit of massive gravity \cite{derham2020}.

\section{Positivity Bounds}\label{sec.bounds}

In this section, we review how the physical assumptions of unitarity, causality, and Lorentz invariance on the UV completion of an EFT gives rise to bounds on the 2-2 scattering amplitudes. We consider four identical particles of mass $m$ and integer spin $S$. The generalization to particles with distinct mass or spin follows a similar derivation and can be found in \cite{de_Rham_2018}, so does the fermionic case. The amplitudes are expressed in terms of the Mandelstam variables $(s,t,u)$ \cite{Mandelstam1958}, where $s$ is the center of mass energy, $t$ the momentum transfer, and $u=4m^2-s-t$ their conjugate variable. In particular, $t$ is related to the scattering angle $\theta$ as
$
    \cos\theta = 1 + \frac{2t}{s-4m^2}
$.
Further definitions of the kinematic variables are presented in Appendix~\ref{sec.kinematics}.

A difficulty that arises in deriving bounds for spinning particles comes from the crossing relations which are trivial in the scalar case, but not in the spinning case (except in the forward limit $t=0$). Therefore, to derive spinning bounds beyond the forward limit, it has been suggested in \cite{de_Rham_2018} to introduce a basis, known as \textit{transversity} basis, which diagonalizes the spinning crossing relations.

The singularities of the 2-2 amplitudes are well-known: there are simple poles at the physical mass in the different exchange channels $s,t,u=m^2$ and a branch cut starting at $s=4m^2$ corresponding to multi-particle production. Crossing symmetry between the $s$ and $u$ channels implies that $A(s,t)$ has two branch cuts on the real $s$-axis, from $s=-\infty$ to $-t$ and $s=4m^2$ to $\infty$, which are referred to as the left hand (LH) and right hand (RH) cuts respectively. The amplitudes are usually assumed to be otherwise analytic in the whole Mandelstam complex plane \cite{Mandelstam1958}. Moreover, the spinning amplitudes have the same domain of analyticity as the scalar ones \cite{MartinSpin,Hara64}. Finally, the relation between causality and analyticity has been established for a long time \cite{Bogoliubov:1959bfo,Bremermann1958}. It is reviewed in the Appendix A of \cite{de_Rham_2018}, and we shall not derive it again.

\paragraph{Tree-level}

The bounds presented in this section are valid to all orders in loops. However, in practice, it turns out to be convenient to work at tree-level. The main difference is that, at tree-level, $\Im[A(s,t)]$ can only be non-zero for $s\geq \Lambda^2$, where $\Lambda$ is the mass of the lightest state outside of the EFT, i.e.\ the cutoff of the EFT \cite{de_Rham_2017}. In this case, we can take the integration along the cuts to start at $\Lambda^2$. In the following, we write the integrals on the branch cuts to run from $\mu_b$ to $\infty$, where $\mu_b=4m^2$, but in general, one can take $\mu_b=\Lambda^2$ at tree-level.

In this work, we compute the bounds at tree-level. This is justified if we assume the presence of a weak coupling in the theory, corresponding to the suppression factor of the loops. As tree and loop effects are mixed in the bounds, it would make no sense to only consider tree-level amplitudes if they did not dominate the loop contributions \cite{de_Rham_2017gal}. We consider weak coupling, but there are recent works exploring beyond it \cite{Bellazzini_2021,Bellazzini2021}.

\subsection{Transversity Formalism}

Here we review the transversity basis introduced in \cite{de_Rham_2018}. The total bosonic\footnote{There is an additional change of signs in the crossing relations for fermions.} amplitude for the process $AB\to CD$ associated with the $s$-channel is related to the corresponding amplitude for $A\bar D\to C\bar B$ (where the bar denotes anti-particles) associated with the $u$-channel by a reordering of the particles \cite{de_Rham_2019},
\begin{equation}
    \mathcal{A}^{s}(p_1,p_2,p_3,p_4) = \mathcal{A}^u(p_1,-p_4,p_3,-p_2)\, .
\end{equation}
The right hand side (RHS) can be expressed in terms of the usual $(p_1,p_2,p_3,p_4)$ configuration via a Lorentz transformation (hence the need for the hypothesis that the UV-theory is Lorentz invariant). Such a Lorentz transformation is trivial for scalar amplitudes, $ A^s(s,t)=A^u(u,t)$, but not for spinning particles.

\begin{figure}
    \centering
    \includegraphics[width = .7\linewidth]{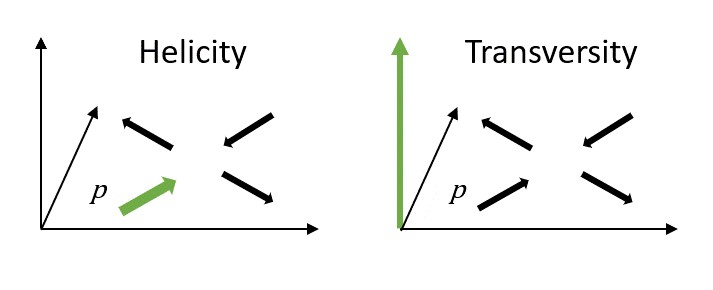}
    \vspace{-1cm}
    \caption{The difference between the helicity and transversity formalism, as introduced in \cite{de_Rham_2018}. The helicity is the spin projection along the direction of motion, whereas the transversity is defined as its projection along the direction transverse to the interaction plane.}
    \label{fig.trans}
\end{figure}

Usually, amplitudes are expressed in helicity basis \cite{JACOB1959404}. However, the crossing relations in this basis are not convenient to deal with, except in the forward limit where positivity bounds for spinning particles have therefore been derived and used in the helicity formalism \cite{Cheung2016,Bellazzini2017,Bonifacio_2016}. Away from the forward limit, they are not diagonal, and not sign-definite \cite{Trueman:1964zzb,Hara:1970gc,Hara71}, such that it is difficult to conclude on the positivity along the LH cut in this formalism, which is a crucial ingredient to derive the positivity bounds. Therefore, to derive bounds beyond the forward limit, we use the transversity formalism \cite{Kotanski66}, which have simplified crossing relations \cite{Kotanski70}.
Helicity polarizations, denoted by $\lambda$, are defined along the momentum direction, while transversity polarizations, denoted by $\tau$, are defined along the direction transverse to the interaction plane. This is depicted in Fig.\ \ref{fig.trans}. Amplitudes in the transverity basis, denoted by $\T$, are related to the ones in helicity basis, denoted by $\H$, by
\begin{equation}\label{eq.htot}
    \mathcal{T}_{\tau_1\tau_2\tau_3\tau_4} = \sum_{\lambda_1\lambda_2\lambda_3\lambda_4}u_{\lambda_1\tau_1}^Su_{\lambda_2\tau_2}^Su_{\lambda_3\tau_3}^{S*}u_{\lambda_4\tau_4}^{S*}\mathcal{H}_{\lambda_1\lambda_2\lambda_3\lambda_4}\, ,
\end{equation}
where $u_{\lambda\tau}^S=D^S_{\lambda\tau}(\pi/2,\pi/2,-\pi/2)$ with $D^S$ a Wigner $D$-matrix \cite{wignergruppentheorie}.
These transversity amplitudes follow nice crossing relations which further simplifiy for elastic scattering or in the forward limit.
\begin{equation}\label{eq.crossrel}
    \mathcal{T}^s_{\tau_1\tau_2\tau_3\tau_4}(s,t,u)=e^{i\pi\sum_i\tau_i}e^{-i\chi_u\sum_i\tau_i}\mathcal{T}^u_{-\tau_1-\tau_2-\tau_3-\tau_4}(u,t,s)\, ,
\end{equation}
with
\begin{equation}\label{eq.chiu}
    e^{\pm i\chi_u}=\frac{-su\mp2im\sqrt{stu}}{\sqrt{s(s-4m^2)u(u-4m^2)}}\, .
\end{equation}

\paragraph{Kinematical singularities}
By its definition \eqref{eq.chiu}, the factor $e^{i\chi_u\sum_i\tau_i}$ in the crossing relation \eqref{eq.crossrel} introduces additional singularities of order $\sum_i\tau_i\leq 4S$. Namely, poles at $s,u=0$, $4m^2$ and a branch point at $\sqrt{stu}=0$. These are kinematical singularities, and it is more convenient to subtract them.

Let us consider these singularities separately. First, it has been shown in \cite{osti_4534874} that helicity amplitudes are regular at $s=0$. Then, by \eqref{eq.htot}, so are the transversity amplitudes. Second, the pole at $s=4m^2$ can be removed by multiplying by a factor $\sqrt{s(s-4m^2)}^{\sum_i\tau_i}$. In practice we use the maximal possible value of the exponent, that is $4S$, so that it works for any configuration of polarizations.
Finally, as $\sqrt{stu}\sim\sin\theta$, we have that $\sqrt{stu}\to-\sqrt{stu}$ under $\theta\to-\theta$, such that any even function of $\theta$ does not contain the branch cut. Altogether, these considerations imply that the quantity
\begin{equation}\label{eq.tplus}
    \mathcal{T}_\taus^+(s,\theta)=\left(s(s-4m^2)\right)^{2S}\left(\mathcal{T}_\taus(s,\theta)+\mathcal{T}_\taus(s,-\theta)\right)\,,
\end{equation}
is free of kinematical singularities. This regularized amplitude plays a similar role to a scalar amplitude in the derivation of the bounds which follows.

\subsection{Spinning Bounds}\label{sec.spinbounds}

\paragraph{Unitarity and analyticity}
The requirement of unitarity imposes, via the optical theorem, that $\Abs_s\T^+_{\tau_1\tau_2}(s,0)$ together with its $t$-derivatives is positive on the RH cut. Thanks to the crossing relations \eqref{eq.crossrel} of the transversity amplitudes, this consideration can be extended to the LH cut.\footnote{This extension is not trivial and can be found in \cite{de_Rham_2018}.} Note that $\Abs_s\T^+(s)=\frac{1}{2i}\text{Disc}\T^+(s)=\frac{1}{2i}\lim_{\epsilon\to 0}[\T^+(s+i\epsilon)-\T^+(s-i\epsilon)]$ denotes the absorptive part of the amplitude, which is equal to the imaginary part if the theory is time reversal invariant. Next, analyticity allows one to analytically continue these properties away from $t=0$, such that
\begin{equation}\label{eq.posabs}
\begin{aligned}
    &\frac{\del^n}{\del t^n} \Abs_s\T^+_{\tau_1\tau_2}(s,t) >0, \quad \forall\, n\geq0, \, s\geq 4m^2, \, 0\leq t<m^2, \\
    &\frac{\del^n}{\del t^n} \Abs_u\T^+_{\tau_1\tau_2}(s,t) >0, \quad \forall\, n\geq0, \, u\geq 4m^2, \, 0\leq t<m^2.
\end{aligned}
\end{equation}

Furthermore, together with unitarity and analyticity, the requirement of locality in a gapped theory implies the presence of a Froissart bound \cite{Froissart,Martin1963}, which also exists for spinning particles \cite{Hara64} and can be extended to non zero $t$ \cite{JinMartin}. With $\epsilon(t)<1$ for $0\leq t< m^2$, it reads
\begin{equation}\label{eq.spinfroissart}
    \lim_{s\to\infty}|\T_\taus(s,t)|< s^{1+\epsilon(t)} \, \implies\, \lim_{s\to\infty}|\T^+_\taus(s,t)|< s^{N_S}.
\end{equation}
The implication for $\T^+(s,t)$ comes from its definition in \eqref{eq.tplus} where it has an additional $s^{4S}$ factor compared to $\T(s,t)$, such that we define
\begin{equation}\label{eq.ns}
    N_S = 4S+2.
\end{equation}
Note that $\T^+$ could have been defined with a minimal factor of $s^{\sum_i\tau_i}$, in which case $N_S=\sum_i\tau_i+2$ is sufficient. However, this definition depending on the configuration of polarizations is not really convenient, especially to study indefinite transversities, so we shall use the definition \eqref{eq.ns} which works for any polarization. It has been argued in \cite{de_Rham_2019} that not using the minimal number of subtractions only leads to small differences and does not change the qualitative form of the bounds.

\paragraph{Dispersion relation}

\begin{figure}[t]
    \centering
    \includegraphics[width=.8\linewidth]{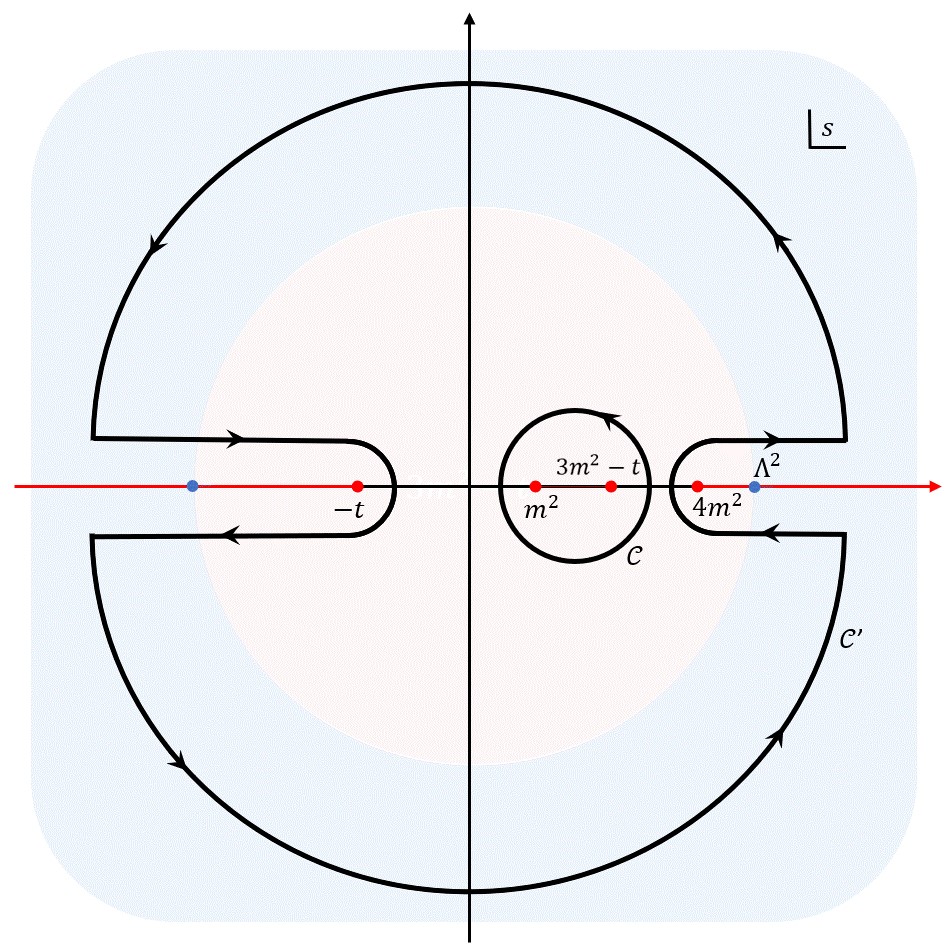}
    \caption{Contours in the complex $s$-plane. The initial contour $\mathcal{C}$ encloses the $s$- and $u$-channel poles situated at $s=m^2$ and $s=3m^2-t$ respectively. The branch cuts, from multi-particle production, are along the real axis $s\geq4m^2$, and by crossing symmetry, $s\leq-t$. The deformed contour $\mathcal{C}'$ goes around them and to $|s|\to\infty$. The red and blue regions represent the regions of validity of the EFT and UV-theory respectively.}
    \label{fig.contour}
\end{figure}

To derive a dispersion relation for the regularized amplitude $\T^+(s,t)$ using Cauchy's theorem, we first define the pole subtracted amplitude
\begin{equation}
    \tilde\T^+_{\tau_1 \tau_2}(s,t)
= \T^+_{\tau_1 \tau_2}(s,t) -  \frac{\text{Res}\T^+_{\tau_1 \tau_2 }(s=m^2,t)}{s-m^2} - \frac{\text{Res}\T^+_{\tau_1 \tau_2} (s=3m^2-t,t)}{s+t-3m^2}\,,
\end{equation}
which is analytic in the whole $s$-complex plane (minus the branch cuts) and can therefore be expressed via a contour integral, see Fig.\ \ref{fig.contour},
\begin{equation}\label{eq.contour}
    \tilde\T^+_{\tau_1 \tau_2}(s,t) = \frac{1}{2\pi i} \oint_\mathcal{C}  ds' \; \frac{\T^+_{\tau_1 \tau_2 }(s',t)}{(s'-s)}\,.
\end{equation}
Next, deforming the contour to $\mathcal{C}'$, the amplitude is given by the arcs at infinity and the contributions along the cuts.
According to \eqref{eq.spinfroissart}, the arc contributions can be dropped by performing $N_S$ subtractions. This introduces subtraction functions $a_n(t)$ and additional powers of $s$. Then, the contributions along the cuts are given by the discontinuity of the amplitude along the cuts; that is the absorptive part. All in all, we can express the contour integral as
\begin{equation}
    \label{eq.disrel1}
    \begin{aligned}
    \tilde\T^+_{\tau_1 \tau_2 }(s,t) =  \sum_{n=0}^{N_S-1}  a_n(t) s^n
&+ \frac{s^{N_S}}{\pi}  \int_{\mu_b}^\infty d \mu  \frac{ {\text{Abs}}_s \T^+_{\tau_1 \tau_2 }(\mu,t) }{ \mu^{N_S} (\mu - s) }   \\&
+\frac{u^{N_S}}{\pi}  \int_{\mu_b}^\infty d \mu  \frac{  {\text{Abs}}_u \T^+_{\tau_1 \tau_2 }(4m^2-t-\mu,t) }{ \mu^{N_S} ( \mu - u) } \,.
    \end{aligned}
\end{equation}

The dispersion relation \eqref{eq.disrel1} allows us to derive positivity constraints on the derivatives of $\tilde\T^+_{\tau_1 \tau_2 }(s,t)$, using \eqref{eq.posabs}. Indeed, defining
\begin{equation}\label{eq.fvtmain}
    f_{\tau_1\tau_2}(v,t)=\frac{1}{N_S!}\frac{\del^{N_S}}{\del s^{N_S}}\tilde\T^+_{\tau_1\tau_2}(s,t)\Big|_{s=v+2m^2-t/2}\, ,
\end{equation}
where the $N_S$ derivatives get rid of the subtraction functions,
the following quantities are positive
\begin{equation}\label{eq.boundssum}
\begin{aligned}
   & \del_v^{2N} f_{\tau_1\tau_2}(v,t)>0 \quad \forall\, N\geq 0,
\\
   &\del_t f_{\tau_1\tau_2}(v,t)+\frac{N_S+1}{2\M^2}f_{\tau_1\tau_2}(v,t)>0 \, ,
\end{aligned}
\end{equation}
where at tree level,
\begin{equation}
    |v|<\Lambda^2, \; 0\leq t<m^2, \; \M^2=\Lambda^2\, .
\end{equation}
In particular, the tree-level bounds can be used to obtain constraints on the EFT cutoff $\Lambda$, as performed in e.g. \cite{de_Rham_2017}. The details of the derivation of \eqref{eq.boundssum} can be found in Appendix~\ref{app.bounds}.

\paragraph{Indefinite transversity bounds}
We can also consider initial and final particles that are not polarized in a definite transversity direction, but rather in some superpositions of transversity polarizations.
The elastic scattering amplitudes for these indefinite transversity states are computed as
\begin{equation}
     \T_{\alpha\beta}(v,t) = \sum_{\taus}\alpha_{\tau_1}\beta_{\tau_2}\alpha^*_{\tau_3}\beta^*_{\tau_4}\T_\taus(v,t)\, ,
\end{equation}
where $\alpha$ and $\beta$ are generic unit vectors. For given indefinite polarizations, their components are the projection of the polarization along the definite transversity polarization vectors. Next, we define the indefinite polarization generalization of $f_{\tau_1\tau_2}$ as
\begin{equation}\label{eq.funpol}
    f_{\alpha\beta}(v,t) = \sum_{\taus}\alpha_{\tau_1}\beta_{\tau_2}\alpha^*_{\tau_3}\beta^*_{\tau_4}f_\taus(v,t)\, ,
\end{equation}
where $f_\taus$ is the inelastic generalization of $f_{\tau_1\tau_2}$, i.e.\ it is computed by \eqref{eq.fvtmain} for $\tilde\T_\taus$.
The arguments we used to prove the positivity of $f_{\tau_1\tau_2}$, and its $v$-derivatives, are still valid for $v=0$ and in the forward limit $t=0$
(see the Appendix A of \cite{de_Rham_2019})\footnote{It is an interesting question whether the condition $v=0$ may be relaxed.}
\begin{equation}
    \del_v^{2N}f_{\alpha\beta}(0,0)>0 \quad \forall \quad N\geq 0\, .
\end{equation}
These bounds hold for indefinite transversities. Therefore, they also hold for definite and indefinite helicities.
Finally, note that it is sufficient to study $f_{\alpha\beta}$ as it contains all the definite polarization quantities: $f_{\tau_1\tau_2}$ corresponds to the $\mathcal{O}(\alpha_{\tau_1}^2\beta_{\tau_2}^2)$ terms.

\section{Bounds on Proca Theories}\label{sec.results}

In the following, we obtain some linear $f$-quantities for the vector theories we study. We can express them as
\begin{equation}\label{eq.fdefmain}
\begin{aligned}
    f_{\tau_1\tau_2}(v,t)&=f_{\tau_1\tau_2}(0,0)+ \del_t f_{\tau_1\tau_2}\cdot t\equiv \frac{1}{\Lambda_2^4}\left[ \mu_{\tau_1\tau_2}+\lambda_{\tau_1\tau_2}\frac{t}{m^2}\right],
\end{aligned}
\end{equation}
with the $\mu$ and $\lambda$ being some combinations of the EFT parameters.
There are four independent quantities for the definite elastic polarizations, denoted
\begin{equation}
    f_{SS}\equiv f_{00}\,,\quad f_{SV}\equiv f_{0\pm1}\, ,\quad f_{V_+}\equiv f_{\pm1\pm1} \, , \quad f_{V_-}\equiv f_{\pm1\mp1} \, .
\end{equation}
We can also write $f_{\alpha\beta}(0,0)$ as
\begin{equation}\label{eq.fabmain}
\begin{aligned}
    f_{\alpha\beta}(0,0)=\frac{1}{\Lambda_2^4}\Big[&\tilde\mu_1|\alpha_+|^2|\beta_+|^2 \\
    +&\tilde\mu_2\big[|\alpha_+|^2(1-|\beta_+|^2)+|\beta_+|^2(1-|\alpha_+|^2)\big] \\
    +&\tilde\mu_3\big[|\alpha_-|^2|\beta_0|^2+|\alpha_0|^2|\beta_-|^2\big] \\
    +&\tilde\mu_4\big[|\alpha_-|^2|\beta_-|^2+|\alpha_0|^2|\beta_0|^2\big]\\
    +&2(\tilde\mu_3-\tilde\mu_4)\Re(\alpha_-\alpha_0^*)\Re(\beta_-\beta_0^*)\\
    +&\tilde\mu_5\big[\Re(\alpha_-\alpha_+^*)\Re(\beta_-\beta_+^*)-\Re(\alpha_0\alpha_+^*)\Re(\beta_0\beta_+^*)\big]\Big]>0\, .
\end{aligned}
\end{equation}
Then, the bounds from the previous section translate into a set of 10 independent bounds:
\begin{equation}\label{eq.sumboundsmain}
\begin{aligned}
    &\tilde\mu_1>0\, , \quad \tilde\mu_2>0 \, ,\quad \tilde\mu_3>0\, , \\
    &\mu_{SS}>0\, ,\quad \mu_{V_+}>0\, , \quad \mu_{V_-}>0\, , \\  &\lambda_{SS}\geq0\, , \quad \lambda_{SV}\geq0\, , \quad \lambda_{V_+}\geq0\, , \quad \lambda_{V_-}\geq0\, .
\end{aligned}
\end{equation}
The formal derivation of this statement is presented in Appendix~\ref{app.linbounds}. The $\mu$'s are bounds in the forward limit; those with a $^\sim$ come from indefinite polarizations. The $\lambda$ bounds correspond to the $t$-derivative bounds, which are available thanks to the analysis beyond the forward limit. We shall now compute these coefficients explicitly for the two types of Proca theories.
\subsection{Generalized Proca}
For convenience, we recall the perturbative interacting Lagrangian \eqref{eq:LGPpert} for Generalized Proca
\begin{equation}\label{eq.lgp2}
\begin{aligned}
    \mathcal{L}_{\text{GP}}^{(3)}&=a_1m^2A^2\del_\mu A^\mu+ a_2\Tilde{F}^{\mu\alpha}\Tilde{F}^\nu_{\  \alpha}\del_\mu A_\nu \\
     \mathcal{L}_{\text{GP}}^{(4)}&=b_1 m^4A^4+b_2m^2A^2F_{\mu\nu}^2+ b_3m^2A^2[(\del\cdot A)^2-\del_\mu A_\nu\del^\nu A^\mu]+b_4m^2A_\mu A^\nu F^{\alpha\mu}F_{\alpha\nu} \\ &\, + b_5F^{\mu\nu}F^{\alpha\beta}F_{\mu\alpha}F_{\nu\beta}+b_6(F_{\mu\nu}^2)^2
    +b_7 \Tilde{F}^{\alpha\beta}\Tilde{F}^{\mu\nu}\del_\alpha A_\mu\del_\beta A_\nu\, .
\end{aligned}
\end{equation}
Computing the amplitudes from this Lagrangian, we obtain $f_{\tau_1\tau_2}$
with the following coefficients, which must be positive,
\begin{equation}\label{eq.gpb1}
\begin{aligned}
    \mu_{SS}=& 8 [2b_5 + 4 b_6+a_2(a_1+\frac{1}{4}a_2)]>0, \qquad
    \mu_{SV}= b_4 + 4 b_5-\frac{1}{2}a_2^2>0, \\
    \mu_{V_+}=& 2[ 2b_5+4 b_6+a_2(a_1+\frac{1}{4}a_2)+a_1^2+b_1-2b_2+b_3]>0,\\
    \mu_{V_-}=&2[2b_5+4 b_6+a_2(a_1-\frac{1}{4}a_2)-3a_1^2+b_1+2b_2-b_3+b_4]>0,
\end{aligned}
\end{equation}
and
\begin{equation}\label{eq.gpb2}
    \begin{aligned}
        \lambda_{SS}=& \frac{3}{2} a_2^2\geq0, \qquad
    \lambda_{SV}= \frac{1}{4}[3a_2^2-4b_7]\geq0,\\
    \lambda_{V_+}=& \frac{3}{2}[b_3+a_1^2+  a_2(a_1+\frac{1}{4}a_2)]\geq0,\\
    \lambda_{V_-}=& \frac{3}{2}[b_3+a_1^2-a_2(a_1+\frac{1}{12}a_2)]\geq0.
    \end{aligned}
\end{equation}
The indefinite polarizations in the forward limit $f_{\alpha\beta}(0,0)$ are of the form given by \eqref{eq.fabmain} with
\begin{equation}\label{eq.gpb3}
    \begin{aligned}
        \tilde\mu_1 &= 8[b_1-a_1^2] >0, \qquad
       \tilde\mu_2 = 2b_4-a_2^2>0, \\
       \tilde\mu_3 &= 8b_5>0,  \qquad
       \tilde\mu_4= 8[2b_5 + 4 b_6+a_2(a_1+\frac{1}{4}a_2)]>0, \\
    \tilde\mu_5 &= -4[4b_2+b_4-2b_3-\frac{1}{2}a_2^2-4a_1^2]\, .
\end{aligned}
\end{equation}
Note that, as mentioned in \eqref{eq.defindef}, the constraint on $\tilde\mu_5$ is encoded in $\mu_{V\pm}$. Additionally, $\mu_{SS}=\tilde\mu_4$ and $\mu_{SV}\propto\tilde\mu_2+\tilde\mu_3$, so some of the previous bounds are redundant.

To summarize, the constraints on the EFT coefficients from the bounds \eqref{eq.sumboundsmain} for Generalized Proca are given by
\begin{equation}\label{eq.gpbounds}
    \begin{aligned}
        & b_1>a_1^2, \quad
        2b_4>a_2^2, \quad
        b_5>0,  \\
        & b_3+a_1^2\geq \text{Max}\left[-a_2(a_1+\frac{1}{4}a_2),a_2(a_1+\frac{1}{12}a_2)\right],\quad
        4b_7\leq 3a_2^2, \\
        &2b_5 + 4 b_6+a_2(a_1+\frac{1}{4}a_2)>0,\\
        & 2b_5+4 b_6+a_2(a_1+\frac{1}{4}a_2)+a_1^2+b_1-2b_2+b_3>0,\\
        &2b_5+4 b_6+a_2(a_1-\frac{1}{4}a_2)-3a_1^2+b_1+2b_2-b_3+b_4>0,
    \end{aligned}
\end{equation}
 where the first line shows the $\tilde\mu$ bounds, the second line the $\lambda$ bounds (without the trivial $\lambda_{SS}$), and the rest are the $\mu$ bounds.
These results are in agreement with those of \cite{de_Rham_2019} (and \cite{Bonifacio_2016}\footnote{Note that the $S$ and $V$ denote helicity polarizations in their results, rather than transversity polarizations.} in the forward limit) up to some parameter redefinitions. We also have additional constraints arising from the interactions parameterized by $a_2$ and $b_7$, which were not previously considered.

Finally, although only $f_{\alpha\beta}(0,0)$ is relevant for the bounds we consider, the full $f_{\alpha\beta}(v,t)$ contains a linear $t$ and $v$ dependence that is worth commenting on. In particular, the $v$ contribution gather all of the $s^3$ terms in the amplitudes (every polarization is included in $f_{\alpha\beta}$) and therefore indicates the scale at which unitarity is perturbatively broken. For the Generalized Proca this contributaion is given by
\begin{equation}\label{eq.fabv}
\begin{aligned}
    f_{\alpha\beta}^v = \frac{14}{\Lambda_2^4m^2}a_2\Big[&a_2\Im(\alpha_-\alpha_0^*)\Im(\beta_0\beta_-^*)\\ &+(2a_1+a_2)[\Im(\alpha_+\alpha_0^*)\Im(\beta_0\beta_+^*)-\Im(\alpha_-\alpha_+^*)\Im(\beta_+\beta_-^*)]\Big].
\end{aligned}
\end{equation}
Then, perturbative unitarity breaks at $s^3\sim\Lambda_2^4m^2\sim\Lambda_3^6$, which confirms the existence of non-trivial operators at the scale $\Lambda_3$.
What is interesting here is the overall $a_2$ coefficient which means that, for any polarizations, the unitarity breaking term is parametrized by $a_2$. Therefore, setting $a_2=0$ raises the cutoff of the model.

\paragraph{Some examples}

\begin{itemize}
\item
First, in addition to raising the cutoff scale, setting $a_2=0$ greatly simplifies the expression of the bounds. It (along with the $b_7$ interaction term) generates scalar-vector mixing term in the decoupling limit \eqref{eq.GPDL}, and, as seen from \eqref{eq.fabv}. The bounds \eqref{eq.gpbounds} for $a_2=0$ are given by
\begin{equation}\label{eq.a20bounds}
    \begin{aligned}
        & b_1>a_1^2, \quad b_4>0, \quad
        b_5>0, \quad
         b_3\geq -a_1^2, \quad b_7\leq 0, \quad 2b_5+4 b_6>0, \\
       &  b_1-2b_2+b_3+a_1^2+2b_5+4 b_6>0,\quad b_1+2b_2-b_3+b_4-3a_1^2 +2b_5+4 b_6>0.
    \end{aligned}
\end{equation}
Hence, having the dimension-6 operator represented by $a_2$ weakens the positivity constraints. This seems to be aligned with the findings in \cite{Zhang:2018shp}.

\item Additionally, $b_7$ only appears in the beyond forward limit bound $\lambda_{SV}\geq 0$. Note here the importance of the beyond forward limit analysis. Without it $b_7$ would remain unconstrained. Therefore, if we consider the case with $b_7=0$ (i.e. canceling the other scalar-vector mixing term than the previous example), we get the bounds \eqref{eq.gpbounds} but without the constraint $4b_7\leq 3a_2^2$ which trivializes. Similarly, setting both $a_2$ and $b_7$ to 0 gives again \eqref{eq.a20bounds}, but without the constraint $b_7\leq 0$.
In other words, the presence or absence of $b_7$ does not influence the constraints on the other parameters.

\item
Then, studying the simplest vector Galileon model, keeping only the interactions in $a_1$ and $b_3$ leads to the inconsistent bounds
\begin{equation}
    3a_1^2<-b_3\leq a_1^2.
\end{equation}
As already noted in \cite{de_Rham_2019}, such a theory admits no standard UV completion. In this work, they indicated that other interactions should be included in order to obtain a window of parameters satisfying the positivity bounds. Here, our analysis on indefinite polarizations provides a more precise result on which interactions to include. The bounds for any model containing the cubic vector Galileon can be satisfied by adding the purely quartic interaction parametrized by $b_1$, which must be positive due to the  condition $b_1>a_1^2$, coming from $\tilde\mu_1$. This minimal model is given by the Lagrangian
\begin{equation}
\begin{aligned}
    \L=&-\frac{1}{4}F^{\mu\nu}F_{\mu\nu}-\frac{1}{2}m^2 A^2 +\frac{1}{\Lambda_2^2}a_1m^2A^2\del_\mu A^\mu \\ &+\frac{1}{\Lambda_2^4} b_3m^2A^2[(\del\cdot A)^2-\del_\mu A_\nu\del^\nu A^\mu] +\frac{1}{\Lambda_2^4} b_1m^4A^4\, ,
\end{aligned}
\end{equation}
with the bounds reducing to
\begin{equation}\label{eq.galbound}
    \begin{aligned}
     3a_1^2-b_1 < -b_3\leq a_1^2.
    \end{aligned}
\end{equation}
The corresponding window of parameter is pictured on figure~\ref{fig.galbounds}.
Note that the bounds on $\tilde\mu_1$ and $\mu_{V_+}$ ($b_1>a_1^2$  and  $b_1+b_3+ a_1^2>0$) are also non-trivial, but are automatically fulfilled when \eqref{eq.galbound} is respected.
\begin{figure}
    \centering
    \includegraphics[width=.78\linewidth]{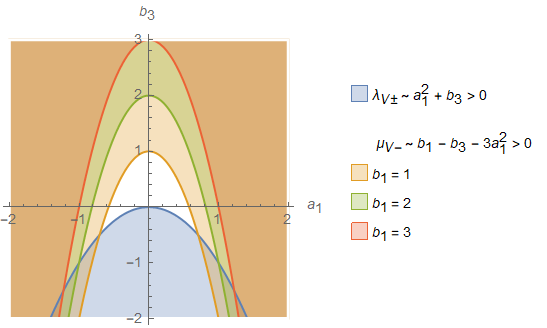}
    \caption{Parameter space of the simplest vector Galileon. The region excluded by $\lambda_{V\pm}$ is indicated in blue. The bound on $\mu_{V_-}$ is pictured for different values of the pure quartic coupling $b_1$. The allowed region increases with $b_1$ and would vanish for $b_1=0$.}
    \label{fig.galbounds}
\end{figure}

\item
If we only consider a model containing the interaction terms leading to scalar-vector mixing in the decoupling limit, with coefficients $a_2$ and $b_7$, then, the bounds cannot be satisfied. The positivity bound, $\tilde\mu_2>0$ requires the $b_4$ interaction to also be included, with the constraints
\begin{equation}\label{eq.ba2b7}
    2b_4>a_2^2, \quad 4b_7\leq 3a_2^2.
\end{equation}
With only these interactions $0\leq\lambda_{V_-}= -a_2^2/8 $ is violated. Hence, we need to also add either $a_1$ (and therefore $b_1$) or $b_3$, i.e.\ again the vector Galileon interactions. The minimal such model would therefore contain $\{a_2,b_3,b_4,b_7\}$ and be constrained by the bounds
\begin{equation}
    \frac{a_2^2}{12}\leq b_3<b_4-\frac{a_2^2}{4}.
\end{equation}

\item
Finally, by the previous considerations, the simplest model including both the vector Galileon $\{a_1, b_3\}$ and mixing interactions $\{a_2,b_7\}$ should also contain the quartic interactions with coefficients $b_1$ (to satisfy $\tilde{\mu}_1>0$) and $b_4$ (to satisfy $\tilde{\mu}_2>0$). The bounds for such a model correspond to setting $b_2,b_5,b_6=0$ in \eqref{eq.gpbounds}. Note that we could also consider only the mixing term $a_2$, setting $b_7=0$, which would give the same bounds.

\item Similarly, the simplest vector Galileon model with $\{a_1, b_3\}$ and only the mixing term $b_7$ should contain $b_1$ as well, and the constraints would be given by \eqref{eq.galbound} plus $b_7\leq 0$.

\end{itemize}

\subsection{Proca-Nuevo}

First, we recall the perturbative Proca-Nuevo Lagrangian, whose coefficients we want to constrain.
\begin{equation}\label{eq.LPNredef}
\begin{aligned}
    &\mathcal{L}_\text{PN}^{(3)}= \gamma_1' m^2 A^2\del_\mu A^\mu+\frac{1}{8}(\alpha_2'-1)[F^2][\del A]+\frac{1}{4}(2-\alpha_2')F^2_{\mu\nu}\del^\mu A^\nu \\
    &\mathcal{L}_\text{PN}^{(4)}= \lambda_0' m^4 A^4 +m^2 A^2\Big[\gamma_2'[\del A]^2-\frac{1}{2}\left(\frac{1}{2}\gamma_1'+\gamma_2'\right)\del_\mu A_\nu \del^\nu A^\mu +\frac{1}{2}\left(\frac{1}{2}\gamma_1'-\gamma_2'\right)\del_\mu A_\nu \del^\mu A^\nu \Big] \\ & \qquad + \frac{1}{128}(-1+\alpha_2'-2\alpha_3')[F^2]^2+\frac{1}{64}(10-5\alpha_2'-14\alpha_3')F^2_{\mu\nu}F^{2\mu\nu} \\ & \qquad+\frac{1}{8}\Big[\alpha_3'[F^2]([\del A]^2-\del_\mu A_\nu\del^\nu A^\mu)+\left(-2+\alpha_2'+4\alpha_3'\right)F^{2\mu\nu}\del^\alpha A_\mu\del_\alpha A_\nu \\& \qquad\qquad +\left(1-\alpha_2'-4\alpha_3'\right)F^{2\mu\nu}[\del A]\del_\mu A_\nu +\left(-2+\alpha_2'+2\alpha_3'\right)F^{\mu\nu}F^{\alpha\beta}\del_\mu A_\alpha\del_\nu A_\beta\Big].
\end{aligned}
\end{equation}
In contrast with the GP model, the PN parameters do not correspond to a specific interaction (except for $\lambda_0'$), but rather appear in various combinations to parametrize several interactions. It is due to the special tuning required to have a ghost-free theory despite higher order EOM. However, note that $\lambda_0'$ parametrizes a purely quartic interaction that vanishes in the decoupling limit, $\gamma_1'$ and $\gamma_2'$ give rise to the cubic and quartic Galileons interactions respectively in the decoupling limit, and the interactions entering with $\alpha$'s give rise to scalar-vector mixing terms in the decoupling limit.

Then, the $f_{\tau_1\tau_2}(v,t)$ is again of the form \eqref{eq.fdefmain} with coefficients that must satisfy the positivity bounds. At constant order the bounds are given by
\begin{equation}\label{eq.pnb1}
\begin{aligned}
    &\mu_{SS}=\frac{1}{8}[2+\alpha_2'(\alpha_2' - 4  (1 + 4 \gamma_1'))]>0, \quad \mu_{SV}=\frac{1}{32}[8 - 2 \alpha_2' -\alpha_2'^2 + 4 \alpha_3']>0, \\
    &\mu_{V_+}=\frac{1}{32}[10+\alpha_2'(\alpha_2'-4(1+4\gamma_1'))-16\gamma_1'(1-4\gamma_1')+96\gamma_2'+64\lambda_0']>0, \\
    &\mu_{V_-}=\frac{1}{32}[2-\alpha_2'(\alpha_2'+4(1+4\gamma_1'))-48\gamma_1'(1+4\gamma_1')-96\gamma_2'+64\lambda_0']>0,
\end{aligned}
\end{equation}
and at linear $t$ order they are given by
\begin{equation}\label{eq.pnb2}
\begin{aligned}
   & \lambda_{SS}=\frac{3}{32}\alpha_2'^2\geq 0, \quad \lambda_{SV}=\frac{1}{64}[3\alpha_2'^2 - 8 \alpha_2' +16 \alpha_3'+4]\geq 0, \\
   & \lambda_{V_+}=\frac{3}{32}[16\gamma_2'+16\gamma_1'^2-\alpha_2'(1+4\gamma_1'-\frac{1}{4}\alpha_2')+2]\geq 0, \\
   & \lambda_{V_-}=\frac{3}{32}[16\gamma_2'+16\gamma_1'^2+\alpha_2'(1+4\gamma_1'-\frac{1}{12}\alpha_2')-\frac{2}{3}]\geq 0.
\end{aligned}
\end{equation}
Additionally, the indefinite polarizations quantity $f_{\alpha\beta}(0,0)$ is given by \eqref{eq.fabmain} with
\begin{equation}\label{eq.pnb3}
\begin{aligned}
    &\tilde\mu_1 = 4[2\lambda_0'-\gamma_1'(1+2\gamma_1')] >0, \quad
   \tilde\mu_2 = \frac{1}{16}[4-\alpha_2'^2]>0, \\
   &\tilde\mu_3 = \frac{1}{8}[2-\alpha_2'+2\alpha_3']>0, \quad
   \tilde\mu_4= \frac{1}{8}[2+\alpha_2'(\alpha_2'-4(1+4\gamma_1'))]>0, \\
    &\tilde\mu_5 = \frac{1}{8}[\alpha_2'^2+16\gamma_1'(1+8\gamma_1')+96\gamma_2'+4].
\end{aligned}
\end{equation}
Before analyzing the bounds, note that the contribution to $f_{\alpha\beta}$ linear in $v$, which corresponds to the $s^3$ terms in all of the amplitudes, is given by
\begin{equation}\label{eq.fabvPN}
\begin{aligned}
    f_{\alpha\beta}^v = \frac{7}{8\Lambda_2^4m^2}\alpha_2'\Big[&\alpha_2'\Im(\alpha_-\alpha_0^*)\Im(\beta_0\beta_-^*)\\ &+(\alpha_2'-8\gamma_1'-2)[\Im(\alpha_+\alpha_0^*)\Im(\beta_0\beta_+^*)-\Im(\alpha_-\alpha_+^*)\Im(\beta_+\beta_-^*)]\Big].
\end{aligned}
\end{equation}
It has exactly the same structure as the GP contribution \eqref{eq.fabv}. Again, this means that perturbative unitarity breaks at scale $\Lambda_3$. The overall coefficient of these terms is now $\alpha_2'$. Therefore, the tuning $\alpha_2'=0$ plays a special role as it raises the cutoff. Unlike for GP, this tuning does not cancel a particular interaction term in the Lagrangian, but rather relates the interactions in a specific way.

\begin{figure}[t!]
\centering
\begin{minipage}{.46\textwidth}
  \centering
  \includegraphics[width=\textwidth]{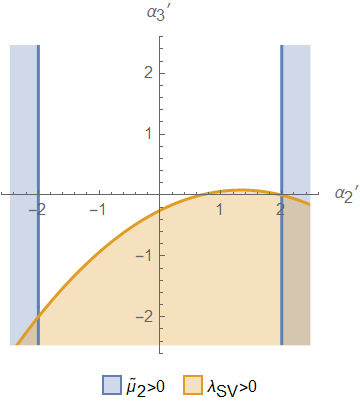}
  \captionof{figure}{Region of the $\{\alpha_2',\alpha_3'\}$ parameter space allowed by the positivity bounds.}
  \label{fig:a3}
\end{minipage}
\quad
\begin{minipage}{.46\textwidth}
  \centering
  \includegraphics[width=\textwidth]{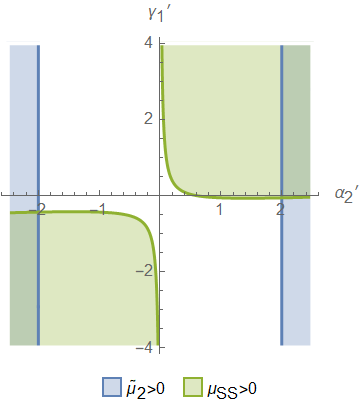}
  \captionof{figure}{Region of the $\{\alpha_2',\gamma_1'\}$ parameter space allowed by $\tilde\mu_2$ and $\mu_{SS}$.}
  \label{fig:c1}
\end{minipage}
\end{figure}

The positivity constraints \eqref{eq.pnb1}$-$\eqref{eq.pnb3} are highly redundant. They reduce to only 5 independent bounds, which are given by
\begin{equation}\label{eq.ReducedBoundsPN}
    \begin{aligned}
         \tilde\mu_2>0 : \quad & -2<\alpha_2'<2\, , \\
         \lambda_{SV}\geq 0 : \quad &\alpha_3'\geq \frac{1}{16}\left(-3\alpha_2'^2+8\alpha_2'-4\right)\, ,\\
         \mu_{SS}>0:  \quad &\alpha_2'\gamma_1'<\frac{1}{16}\left(\alpha_2'^2-4\alpha_2'+2\right)\, ,\\
        \lambda_{V_-}\geq 0 : \quad & \gamma_2'\geq -\gamma_1'^2+\frac{1}{192}\alpha_2'(\alpha_2'-48\gamma_1'-12)+\frac{1}{24}\, ,\\
         \mu_{V_-}>0 : \quad & \lambda_0'-\frac{3}{2}\gamma_2'>3\gamma_1'^2+\frac{3}{4}\gamma_1'+\frac{1}{64}\alpha_2'(\alpha_2'+16\gamma_1'+4)-\frac{1}{32}\, .
    \end{aligned}
\end{equation}
Both the analysis of indefinite polarizations, via $\tilde\mu_2$, and beyond forward limit, via $\lambda_{SV}$ and $\lambda_{V_-}$, play an important role in constraining the model. In the following, we comment on the constraints \eqref{eq.ReducedBoundsPN} separately and show them graphically on figures~\ref{fig:a3}$-$\ref{fig.grid}, where the region excluded by each bound is depicted in its own color and the allowed region remains white.

First, we note that the particular case $\alpha_2'=0$ greatly simplifies the bounds.
Even if this is not necessarily manifest from the Lagrangians, $\alpha_2'$ appears to play a similar role to $a_2$ from the GP model, both in the bounds and in the breaking of perturbative unitarity. Setting $\alpha_2'=0$, the bounds on $\tilde\mu_2$ and $\lambda_{SV}$ are trivially respected, and the other ones simplify to
\begin{equation}\label{eq.PNBa20}
    \begin{aligned}
         \lambda_{SV}\geq 0 : \quad &\alpha_3'\geq -\frac{1}{4}\, ,\\
        \lambda_{V_-}\geq0 : \quad & \gamma_2'\geq-\gamma_1'^2+\frac{1}{24}\, ,\\
         \mu_{V_-}>0 : \quad & \lambda_0'-\frac{3}{2}\gamma_2'>3\gamma_1'^2+\frac{3}{4}\gamma_1-\frac{1}{32}\, .
    \end{aligned}
\end{equation}

\afterpage{
\begin{figure}[ht!]
    \centering
    \begin{subfigure}[c]{0.48\textwidth}
        \includegraphics[width=\textwidth]{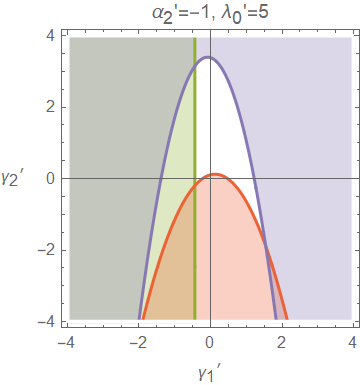} \\
    \end{subfigure}
    \hfill
    \begin{subfigure}[c]{0.49\textwidth}
        \vspace{.2cm}
        \includegraphics[width=\textwidth]{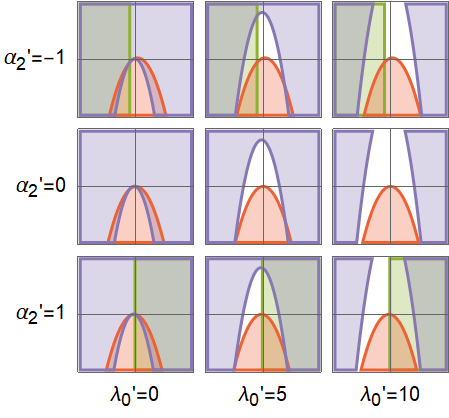} \\
    \end{subfigure}
    \begin{subfigure}[c]{0.48\textwidth}
        \includegraphics[width=\textwidth]{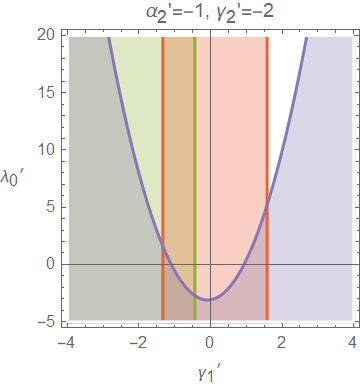}
    \end{subfigure}
    \hfill
    \begin{subfigure}[c]{0.49\textwidth}
        \vspace{.1cm}
        \includegraphics[width=\textwidth]{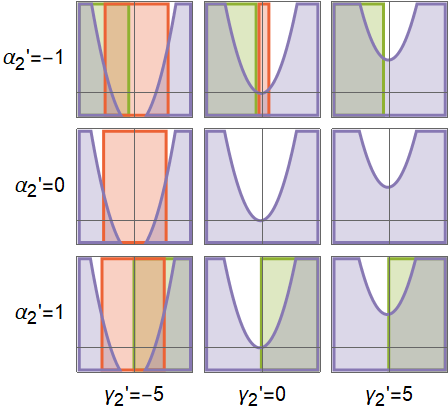}
    \end{subfigure}
    \includegraphics[width=.37\linewidth]{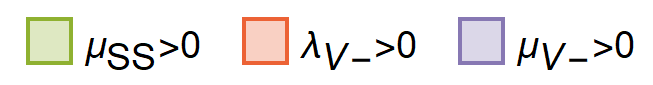}
    \caption{Regions of the parameter space excluded by the three bounds relating $\{\alpha_2',\gamma_1',\gamma_2',\lambda_0'\}$. \textbf{Upper}: This plots the $\{\gamma_1',\gamma_2'\}$ plane. \textbf{Lower}: This plots the $\{\gamma_1',\lambda_0'\}$ plane. \textbf{Left}: This is for a given choice of parameters, allowing for a better visualisation and showing the axes ranges. \textbf{Right} : This is for various choice of the parameters. The omitted axes are the same as those on the left. The regions allowed by all of the bounds remain white.}
    \label{fig.grid}
\end{figure}
\clearpage}

Now, back to the general $\alpha_2'$ case, the first bound of \eqref{eq.ReducedBoundsPN}, $\tilde\mu_2>0$ restricts one of the parameter, $\alpha_2'$, in a $\mathcal{O}(1)$ interval, independently of the rest of the model. In particular, the coefficients of the interactions $[F^2][\del A]$ and $F^2_{\mu\nu}\del^\mu A^\nu$ in $\L_\text{PN}^{(3)}$ are constrained to lie in the intervals $(-3/8,1/8)$ and $(0,1)$ respectively.

Then, $\alpha_3'$ only appears in the second bound, $\lambda_{SV}>0$, together with $\alpha_2'$. The allowed region for $\{\alpha_2',\alpha_3'\}$ is therefore entirely determined by the first two bounds and is shown in figure~\ref{fig:a3}. Although neither of these coefficients corresponds to a specific interaction term by itself, together they parametrize the last three lines of $\L_\text{PN}^{(4)}$, as can be seen in \eqref{eq.LPNredef}.

Next, $\mu_{SS}>0$ relates $\gamma_1'$ to $\alpha_2'$ as shown in figure~\ref{fig:c1}. This region corresponds to
\begin{equation}\label{eq.signc1}
    \begin{aligned}
    &\begin{cases}
    \gamma_1'> F(\alpha_2')\quad &-2<\alpha_2'<0, \\ \gamma_1'\in\mathbb{R} &\alpha_2'=0, \\
    \gamma_1'< F(\alpha_2')\quad &0<\alpha_2'<2,
    \end{cases}
    \qquad F(\alpha_2')=\frac{1}{16}\frac{\alpha_2'^2-4\alpha_2'+2}{\alpha_2'}\,.
    \end{aligned}
\end{equation}
However, the coefficient $\gamma_1'$ also appears in the remaining bounds. Adding these constraints in figure~\ref{fig:c1} would reduce the allowed region.

\begin{figure}[t!]
    \centering
    \begin{subfigure}[c]{0.67\textwidth}
        \includegraphics[width=\textwidth]{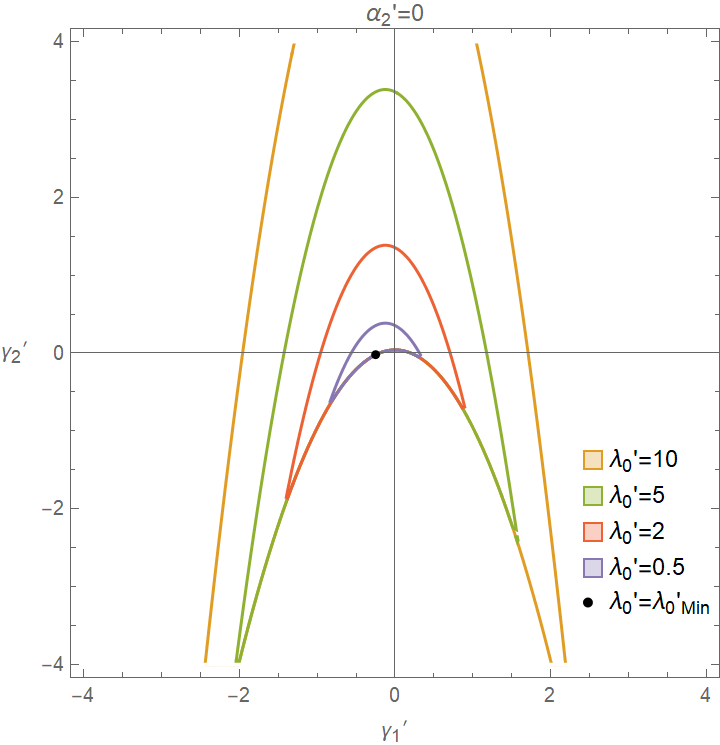}
    \end{subfigure}
    \begin{subfigure}[c]{0.288\textwidth}
    \vspace{-.85cm}
        \includegraphics[width=\textwidth]{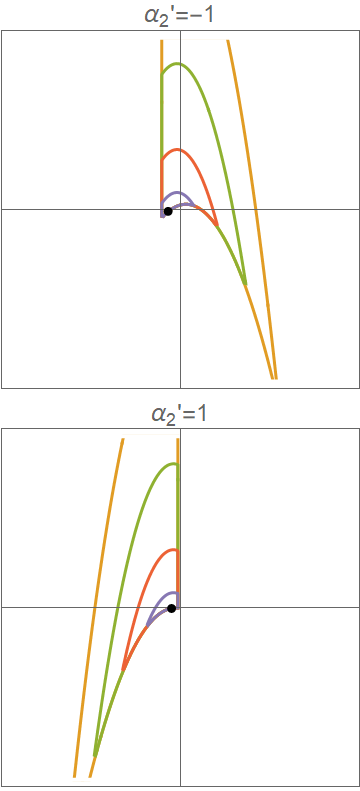}
    \end{subfigure}
    \caption{Allowed islands of the $\{\gamma_1',\gamma_2'\}$ parameter space for various values of $\lambda_0'$ and $\alpha_2'$. The regions shrink for decreasing $\lambda_0'$, to eventually become a point at $\lambda_0'=\lambda'_{0\text{ Min}}$ and vanish for smaller values. For $\alpha_2'=0,-1,1$, this happens at $\lambda'_{0\text{ Min}}\approx-0.062,-0.042,-0.042$ at the points $(\gamma_1',\gamma_2')\approx(-0.25, -0.02),(-0.29, -0.04),(-0.29, -0.01)$ respectively.}
    \label{fig.PNgamma2}
\end{figure}

Finally, $\gamma_2'$ and $\lambda_0'$ are successively introduced in $\lambda_{V_-}$ and $\mu_{V_-}$. These constraints are considered together with those on $\alpha_2'$ and $\gamma_1$. These bounds are shown both in the $\{\gamma_1',\gamma_2'\}$ and $\{\gamma_1',\lambda_0'\}$ planes for various values of the other parameters in the top and bottom lines of figure~\ref{fig.grid} respectively. These allow for a visualization of the influence of different parameters on the allowed regions of the parameter space, shown in white. In particular, we pick $\alpha_2'$ to be successively negative, zero, and positive. Regarding the top line, the islands are reproduced in figure~\ref{fig.PNgamma2} for various values of $\lambda_0'$. We observe that the islands shrink for decreasing values of $\lambda_0'$, with no allowed region for $\lambda_0'< -0.0624$. This bound on $\lambda_0'$, which parametrizes the quartic interaction $A^4$ is similar to the bound on $b_1$ in GP, which also parametrizes the $A^4$ interaction. However, in GP, $b_1$ must be positive, while $\lambda_0'$ is still allowed to be very slightly negative.
From the bottom line of figure~\ref{fig.grid}, we can see that the allowed region of the $\{\gamma_1',\lambda_0'\}$ parameter space is the parabola described by $\mu_{V_-}$, which gets shifted upwards for increasing $\gamma_2'$. Also, a band around $\gamma_1'=0$ is forbidden by $\lambda_{V_-}$ for negative $\gamma_2'$. In particular, the region $\lambda_0'< -0.0624$ is always forbidden by one of the bounds, in accordance with our previous remarks.

\begin{figure}[ht!]
    \centering
    \includegraphics[width=.39\linewidth]{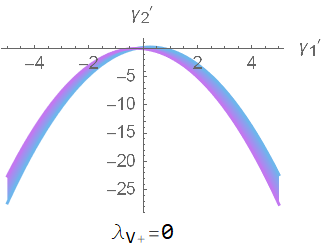} \quad
    \includegraphics[width=.39\linewidth]{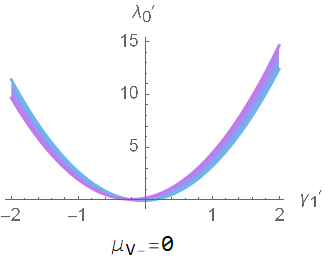}\quad
    \includegraphics[width=.15\linewidth]{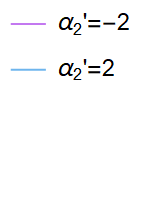}
    \caption{Boundaries of the $\lambda_{V_-}$ and $\mu_{V_-}$ bounds for the allowed interval $-2<\alpha_2'<2$. The curves stay close to one another for the different values of $\alpha_2'$, indicating that $\alpha_2'$ has only little influence on these bounds. In the right plot, we set $\gamma'_2=0$, as other values of $\gamma_2$ would only shift the plot.}
    \label{fig.a2influence}
\end{figure}
\begin{figure}[ht!]
    \centering
    \includegraphics[width=.39\linewidth]{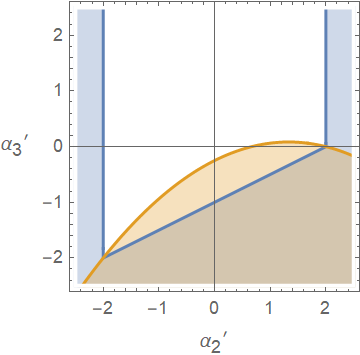} \quad
    \includegraphics[width=.39\linewidth]{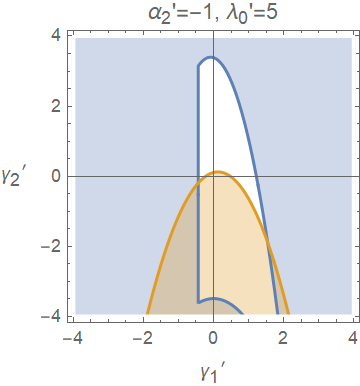}\quad
    \includegraphics[width=.15\linewidth]{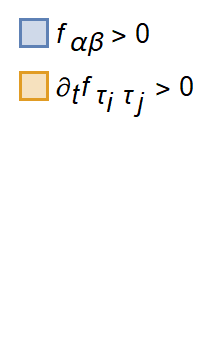}
    \caption{Regions excluded by the $\ptwiddle{\mu}$
    bounds in the forward limit (blue) and the $\lambda$ bounds beyond (orange) in the $\{\alpha_2',\alpha_3'\}$ and  $\{\gamma_1',\gamma_2'\}$ planes (for an arbitrary value of $\alpha_2'$ and $\lambda_0'$). The final white regions respectively correspond to the ones on figure~\ref{fig:a3} and figure~\ref{fig.grid} up left.}
    \label{fig.bflinfluence}
\end{figure}
The value of $\alpha_2'$ primarily influences the green region in figure~\ref{fig.grid} and the blue and red regions are fairly stable as $\alpha_2'$ changes.
This can be seen explicitly by inspecting the last two bounds for various values of $\alpha_2'$ in its allowed interval, see figure~\ref{fig.a2influence}. Therefore, the $\lambda_{V_-}$ and $\mu_{V_-}$ bounds, whose expressions may seem rather complicated at first glance on \eqref{eq.ReducedBoundsPN}, are well approximated by their simplified expressions at $\alpha_2'=0$ in \eqref{eq.PNBa20}. Accordingly, the main role of $\alpha_2'$ in the three last bounds of \eqref{eq.ReducedBoundsPN} is roughly to pick the sign of $\gamma_1'$ via \eqref{eq.signc1}. This can be seen in figure~\ref{fig.PNgamma2} where only the left/right part of the full island is allowed for positive/negative $\alpha_2'$, but the rest of the shape remains quite stable.

As a last comment, we investigate the importance of the beyond forward limit analysis. It was observed in \cite{de_Rham_2019} that the bounds beyond the forward limit had only little influence on the allowed island for dRGT massive gravity and it is interesting to notice that the same does not hold for PN. Inspecting figure~\ref{fig.bflinfluence}, we see that going away from forward limit is particularly  relevant for PN. In particular, the island for $\{\gamma_1',\gamma_2'\}$ is reduced by more than a half. However, we should note that we have not used the most stringent bounds in the forward limit, and by minimizing $f_{\alpha\beta}$ numerically, it is possible that the blue region could be closer to the orange region. In either case, this illustrates how, when applied on the decoupling limit of a theory, the impact of positivity bounds may significant differ as compared to the original theory. In itself this statement is not surprising as noted in \cite{de_Rham_2017gal} as taking the decoupling limit of a theory is a different procedure than considering the low-energy EFT and the mixing with other IR operators may be significant. 

\subsection{Comparison between GP and PN}\label{sec.comparison}

Even though there exists no tuning of the coefficients that makes GP and PN equivalent, we find some of the coefficients play  similar roles in both theories. This identification is rendered more straightforward by our redefinition \eqref{eq.PNredef} of the PN coefficients. In particular, from our previous results, there is an obvious analogy between $a_2$ and $\alpha_2'$, although they play a quite different at the Lagrangian level, as they both control the scale at which perturbative unitarity breaks in their respective theories. Also, the pure quartic interaction $A^4$ is parametrized by $b_1$ and $\lambda_0'$ in each model. Finally, by looking at the decoupling limits \eqref{eq.GPDL} and \eqref{eq.PNDL}, we see that $a_1$ and $b_3$ are analogous to $\gamma_1'$ and $\gamma_2'$, in the sense that they correspond to the cubic and quartic Galileon interactions respectively. However, they are not truly equivalent parameters, as $\gamma_1'$ and $\gamma_2'$ appear also as the coefficients of additional interactions away from the decoupling limit. To summarize, this analogy of coefficients is given by
\begin{equation}\label{eq.param}
    \begin{aligned}
        &a_1\,\longleftrightarrow\,\gamma_1'\, , \quad a_2\,\longleftrightarrow\,\alpha_2'\, , \quad b_1\,\longleftrightarrow\,\lambda_0' \, , \quad b_3\,\longleftrightarrow\,\gamma_2' \, .
    \end{aligned}
\end{equation}

The results for the ten bounds \eqref{eq.sumboundsmain} for GP and PN are reported in table~\ref{tab.bounds}.
Inspecting these results, while keeping this analogy in mind, we observe a similarity between the bounds.
\begin{table}[h!]
\renewcommand{\arraystretch}{1.5}
    \centering
    \begin{tabular}{|l|l|l|} \hline
    & Generalized Proca & Proca-Nuevo  \\ \hline
        $\tilde\mu_1>0$ & $b_1-a_1^2>0$& $2\lambda_0'-\gamma_1'(1+2\gamma_1')>0$ \\
        $\tilde\mu_2>0$  & $2b_4-a_2^2>0$ &$4-\alpha_2'^2>0$ \\
        $\tilde\mu_3>0$  & $b_5>0$& $2-\alpha_2'+2\alpha_3'>0$\\
        $\mu_{SS}>0$   & $2b_5+4 b_6+a_2(a_1+\frac{1}{4}a_2)>0$ &$2+\alpha_2'(\alpha_2'-4(1+4\gamma_1'))>0 $ \\
        $\mu_{V_+}>0$ &   $2b_5+4 b_6+a_2(a_1+\frac{1}{4}a_2)$&$10+\alpha_2'(\alpha_2'-4(1+4\gamma_1'))$ \\&$\quad+a_1^2+b_1-2b_2+b_3>0$ & $\quad -16\gamma_1'(1-4\gamma_1')+96\gamma_2'+64\lambda_0'>0$ \\
         $\mu_{V_-}>0$&$2b_5+4 b_6+a_2(a_1-\frac{1}{4}a_2)$&$ 2-\alpha_2'(\alpha_2'+4(1+4\gamma_1'))$\\
         &$\quad-3a_1^2+b_1+2b_2-b_3+b_4>0$&$\quad -48\gamma_1'(1+4\gamma_1')-96\gamma_2'+64\lambda_0'>0 $ \\
        $\lambda_{SS}\geq0$&$a_2^2\geq0$  &$\alpha_2'^2\geq0$\\
        $\lambda_{SV}\geq0$   &$a_2^2-4b_7\geq0$ &$ 3\alpha_2'^2 - 8 \alpha_2' +16 \alpha_3'+4\geq0$\\
        $\lambda_{V_+}\geq0$    & $b_3 + a_1^2+a_2(a_1+\frac{1}{4}a_2)\geq0$ &$16\gamma_2'+16\gamma_1'^2-\alpha_2'(1+4\gamma_1'-\frac{1}{4}\alpha_2')+2\geq0$\\
        $\lambda_{V_-}\geq0$  &$ b_3  +a_1^2-a_2(a_1+\frac{1}{12}a_2)\geq0$&$16\gamma_2'+16\gamma_1'^2+\alpha_2'(1+4\gamma_1'-\frac{1}{12}\alpha_2')-\frac{2}{3}\geq0$\\ \hline
    \end{tabular}
    \caption{Summary of the positivity bounds \eqref{eq.gpb1}$-$\eqref{eq.gpb3} for GP and \eqref{eq.pnb1}$-$\eqref{eq.pnb3} for PN. The positive overall factors are omitted.
    The coefficients of the GP and PN models are defined in \eqref{eq:LGPpert} and \eqref{eq.LPNredef1} respectively.}
    \label{tab.bounds}
    \end{table}

Moreover, re-establishing the overall coefficients that were omitted in table~\ref{tab.bounds} for readability, there exists a tuning that makes the bounds of both theories perfectly equivalent, which is given by
\begin{equation}\label{eq.boundstuning}
    \begin{aligned}
        &a_1=\pm(\gamma_1'+\frac{1}{4}-\frac{1}{3}\alpha_2'^{-1}) \\
&a_2=\pm\frac{1}{4}\alpha_2' \\
&b_1=\lambda_0'+\frac{1}{16}-\frac{1}{6}\alpha_2'^{-1}(1+4\gamma_1')+\frac{1}{9}\alpha_2'^{-2} \\
&b_3=\gamma_2'-\frac{1}{2}\gamma_1'-\frac{1}{48}+\frac{1}{6}\alpha_2'^{-1}(1+4\gamma_1')-\frac{1}{9}\alpha_2'^{-2}\\
&b_7=-\frac{1}{16}(1-2\alpha_2'+4\alpha_3')
\end{aligned} \qquad
\begin{aligned}
&b_2=-\frac{1}{4}\gamma_2'+\frac{1}{8}\gamma_1'-\frac{1}{96} \\ &\qquad-\frac{1}{12}\alpha_2'^{-1}(1+4\gamma_1')+\frac{1}{8}\alpha_2'^{-2} \\
&b_4=\frac{1}{8} \\
&b_5=\frac{1}{64}(2-\alpha_2'+2\alpha_3') \\
&b_6=-\frac{1}{384}(11-3\alpha_2'+6\alpha_3')\, .
    \end{aligned}
\end{equation}
The first terms correspond to the instinctive matching of \eqref{eq.param}, but there are additional corrections depending mainly on $\alpha_2'$. We could imagine that the tuning takes a simpler form if we consider the bounds with $\alpha_2'=0$. However, quite surprisingly, there is no such tuning in this case. The existence of this matching does not mean that the theories are equivalent as these bounds only contain the $s^2$ contributions to only the elastic amplitudes. Inserting this solution in the full amplitudes does not make them match. For instance, considering the same amplitude in helicity basis as \cite{derham2020},
\begin{equation}
    \mathcal{A}_{++--}^{\text{GP}}(s,t)-\mathcal{A}_{++--}^{\text{PN}}(s,t)=\frac{1}{96\Lambda_2^4}\, s \left(3 \cos (2 \theta ) \left(s-4 m^2\right)+44 m^2-3 s\right).
\end{equation}
Therefore, there is no profound meaning to the solution of \eqref{eq.boundstuning}, it is simply a tuning that makes the bounds equivalent.

Another consequence of the similarity of the bounds is that the study of the simplest Galileon from GP with parameters $\{a_1,b_3,b_1\}$ relates very well with the $\{\gamma_1',\gamma_2',\lambda_0'\}$ sector of PN for $a_2=\alpha_2'=0$. Indeed, keeping in mind the analogy \eqref{eq.param} and reporting these bounds from \eqref{eq.galbound} and \eqref{eq.PNBa20} for GP and PN respectively
\begin{equation}
    \begin{aligned}
        \lambda_{V_-}\geq0 &\, : \, b_3\geq-a_1^2, \quad\qquad  \gamma_2'\geq-\gamma_1'^2+\frac{1}{24}\\
        \mu_{V_-}>0 &\, : \,  b_1-b_3>3a_1^2, \quad  \, \lambda_0'-\frac{3}{2}\gamma_2'>3\gamma_1'^2+\frac{3}{4}\gamma_1'-\frac{1}{32},
    \end{aligned}
\end{equation}
we observe a similar structure. It can also be seen by comparing figure~\ref{fig.galbounds} with the left of figure~\ref{fig.PNgamma2}. Note that, even in this case, PN still has a full additional sector parametrized by $\alpha_3'$ that can not be removed.

Finally, we found in both models that the quartic interaction $A^4$ is bounded from below, needing to be positive for GP ($b_1>0$) and for PN, $\lambda_0'>-0.06$. Possibly, some more sringent bounds on PN would constrain $\lambda_0'$ to be strictly positive as well.

\section{Conclusion}
In this work, we have reviewed how EFT amplitudes for spinning particles are constrained  by assuming a unitary, causal, Lorentz invariant UV completion and how this motivates the use of the transversity formalism due to its crossing symmetry properties to derive positivity bounds away from the forward limit.
The application of these bounds to the Generalized Proca and Proca-Nuevo has allowed us to strongly restrict the parameter space of these models. We have first studied them separately, by exhibiting a set of ten inequalities each of their coefficients had to satisfy. This set further reduced to five independent bounds for PN, confining its parameters in islands, which we displayed in several figures.
In particular, this work furnishes the first positivity bounds analysis of the PN model.
Finally, we have highlighted an analogy between certain coefficients between the two theories, which lead to similar structures in the bounds for both theories.
Overall, we have found that PN is more constrained by the bounds than GP.
This makes sense as the PN coefficients are associated with multiple interactions, due to the non-linear realization of the Hessian constraint, and therefore appear repetitively in the bounds. We emphasize that this analysis has been performed at tree-level however the one-loop corrections considered in \cite{Zosso2021,derham2021quantum} where shown to arise at a higher scale.

From a technical point of view, there would be various ways to tighten the bounds we derived in this work. First, we could use insights from fully triple crossing symmetric bounds as derived in \cite{Sinha:2020win,Haldar:2021rri,Raman:2021pkf,Chowdhury:2021ynh,Sinha:2022sdo,tolley2021new,Caron-Huot2021,Du_2021}, however applying those bounds beyond the forward limit will require further generalizing the formalism.
Second, the minimization of the indefinite polarization bound \eqref{eq.fabmain} could be made more precise, either using some numerical methods \cite{Cheung2016} or via a more systematic analytical minimization \cite{Zhou2019,Zhou2021}.  These bounds could also potentially be further tighten or complemented with the use of pure causality bounds as illustrated in \cite{CarrilloGonzalez:2022fwg}. The application of these causality bounds could prove particularly useful when considering these EFTs on curved background as would be relevant for cosmology and Black Hole constraints.

In practice, the bounds we derived could be used to restrict the parameter space of dark energy models, in complement to observational data.
This analysis could also be combined with other constraints on the parameters arising for example from the presence of a Vainshtein mechanism on spherically symmetric background, as studied in \cite{De_Felice_2016} for the Generalized Proca, or else from imposing subluminal propagation of gravitational waves \cite{Melville_2020,Noller2021}.
Similarly, these bounds could be joined with restrictions coming from stability analysis of quantum corrections or Swampland conjectures. 

Finally, it may be interesting to further study the perturbative PN model \eqref{eq.LPNredef1} under the redefinition of parameters we have proposed in \eqref{eq.PNredef}. In particular, it eliminates a free parameter.
Moreover, the tuning $\alpha_2'=0$, which does not obviously play a particular role at the Lagrangian level,
seems to give an interesting realisation of the model, as we have notably shown in \eqref{eq.fabvPN} that it raises the cutoff.

Even though GP and PN are fundamentally different theories, if one restricts their parameter space adequately they might share equivalent positivity bounds and suffer from similar quantum corrections, which might then indicate that they descent from the same more fundamental UV-complete theory. As was shown in \cite{Zosso2021,derham2021quantum}, both have the same high energy behaviour when quantum corrections are included. Loop induced counter terms have the exact same structure and scaling.
In this work we have further shown, that despite of the fact that PN is more constrained by the positivity bounds than GP, similar structures in the bounds for both theories arise and this might signal that their spin-1 field originates from the same UV theory and the general properties of this fundamental theory give rise to the same structures.

As a final note, it is worth emphasizing that the results presented here are relevant for beyond models of dark energy. In constraining allowed massive spin-1 interactions, this framework opens the door for better understanding of the allowed operators for vector bosons beyond the standard model and could be relevant for models of dark matter, particularly those involving a (massive) dark photon.

\newpage
\appendix

\section{Set-up} \label{sec.kinematics}
We consider 2-2 amplitudes of four identical particles of mass $m$ and spin $S$ with momenta $p_1$, $p_2$, $p_3$ and $p_4$. Working in the center of mass frame, considering the scattering in the $xz$-plane, and treating the particles as all incoming by flipping the sign of the outgoing particles' momenta, the momenta are given by
\begin{equation}\label{eq.mom}
    p_i = (-1)^\eta(E,p\cos\theta_i,0,p\sin\theta_i)\,,
\end{equation}
with $\eta=0$ for the incoming momenta $p_1$ and $p_2$, $\eta=1$ for the outgoing momenta $p_3$ and $p_4$, and
\begin{equation}
    \theta_1 = 0, \quad \theta_2 = \pi \quad \theta_3 = \theta, \quad \theta_4 = \pi+\theta.
\end{equation}
We work with the Mandelstam variables $s=-(p_1+p_2)^2, t=-(p_1+p_3)^2$, and $u=-(p_1+p_4)^2$, with $s+t+u=m^2$.
The parameters of \eqref{eq.mom} can be expressed in terms of the Mandelstam variables as
\begin{equation}\label{eq.eptheta}
    E = \frac{\sqrt{s}}{2}, \quad p = \frac{1}{2}\sqrt{s-4m^2}, \quad \cos\theta = 1 + \frac{2t}{s-4m^2}, \quad \sin\theta=2\frac{\sqrt{tu}}{(s-4m^2)},
\end{equation}
such that the physical region corresponds to $s>4m^2$.
Any two independent variables are sufficient to express the theory. Examples of such sets are $(p, \theta)$ or else $(s,t)$. Also, note that the forward limit $t=0$ corresponds to $\cos\theta=1$, or equivalently, $\theta=0$, hence the name forward.

To compute amplitudes in transversity basis, we use the polarization vectors from \cite{de_Rham_2018}, given by
\begin{equation}
    \begin{aligned}
        &\epsilon^\mu_{\tau=\pm1}(p_i)=\frac{i}{\sqrt{2}m}(p,E\sin\theta_i\pm im\cos\theta_i,0,E\cos\theta_i\mp im\sin\theta_i), \\
        &\epsilon^\mu_{\tau=0}(p_i)=(0,0,1,0).
    \end{aligned}
\end{equation}
\section{Bounds Derivation}\label{app.bounds}
First, we define a quantity that gets rid off the subtraction functions by taking $N_S$ derivatives on $\tilde\T^+$, defined in \eqref{eq.tplus}, as follows
\begin{equation}
    f_{\tau_1\tau_2}(s,t)=\frac{1}{N_S!}\frac{\del^{N_S}}{\del s^{N_S}}\tilde\T^+_{\tau_1\tau_2}(s,t)\, .
\end{equation}
Using the dispersion relation \eqref{eq.disrel1}, or equivalently taking the derivatives on the contour integral \eqref{eq.contour} and deforming the contour afterwards, we can write
\begin{equation}
    f_{\tau_1\tau_2}(s,t)=\frac{1}{\pi}\int_{\mu_b}^\infty d\mu \frac{ {\text{Abs}}_s \T^+_{\tau_1 \tau_2 }(\mu,t) }{ (\mu - s)^{N_S+1} } +\frac{1}{\pi}\int_{\mu_b}^\infty d\mu  \frac{  {\text{Abs}}_u \T^+_{\tau_1 \tau_2 }(4m^2-t-\mu,t) }{ ( \mu - u)^{N_S+1} }\, .
\end{equation}
We know from \eqref{eq.posabs} that the numerators are positive for $0\leq t<m^2$. The denominators are also positive for $s$ within the branch cuts, such that
\begin{equation}
    f_{\tau_1\tau_2}(s,t)>0 \quad \text{for } \;  4m^2-t-\mu_b<s<\mu_b, \; 0\leq t<m^2.
\end{equation}

Now, let us consider this quantity in terms of the crossing symmetric variable $v=s+t/2-2m^2$,
\begin{equation}\label{eq.fvt}
    f_{\tau_1\tau_2}(v,t)=f_{\tau_1\tau_2}(s=v+2m^2-t/2,t)=\frac{1}{N_S!}\frac{\del^{N_S}}{\del s^{N_S}}\tilde\T^+_{\tau_1\tau_2}(s,t)\Big|_{s=v+2m^2-t/2}\, .
\end{equation}
Then,
\begin{equation}
    f_{\tau_1\tau_2}(v,t)=\frac{1}{\pi}\int_{\mu_b}^\infty d\mu \frac{ {\text{Abs}}_s \T^+_{\tau_1 \tau_2 }(\mu,t) }{ (\mu - 2m^2+t/2-v)^{N_S+1} } + \frac{  {\text{Abs}}_u \T^+_{\tau_1 \tau_2 }(4m^2-t-\mu,t) }{ ( \mu -2m^2+t/2+v)^{N_S+1} }\, ,
\end{equation}
and we still have
\begin{equation}
    f_{\tau_1\tau_2}(v,t)>0 \quad \text{for} \;  |v|<\mu_b-2m^2+t/2, \, 0\leq t<m^2.
\end{equation}
Furthermore, the $v$-derivatives are given by
\begin{equation}
\begin{aligned}
    \del_v^N f_{\tau_1\tau_2}(v,t)=\frac{(N_S+N)!}{N!}\frac{1}{\pi}\int_{\mu_b}^\infty d\mu &\frac{ {\text{Abs}}_s \T^+_{\tau_1 \tau_2 }(\mu,t) }{ (\mu - 2m^2+t/2-v)^{N_S+1+N} }\\ +(-1)^N &\frac{  {\text{Abs}}_u \T^+_{\tau_1 \tau_2 }(4m^2-t-\mu,t) }{ ( \mu -2m^2+t/2+v)^{N_S+1+N} }\, ,
\end{aligned}
\end{equation}
such that every even $v$-derivative is positive
\begin{equation}
    \del_v^{2N} f_{\tau_1\tau_2}(v,t)>0 \quad \forall\, N\geq 0, \quad  |v|<\mu_b-2m^2+t/2, \, 0\leq t<m^2.
\end{equation}
Next, looking at the first $t$-derivative,
\begin{equation}
\begin{aligned}
    \del_t f_{\tau_1\tau_2}(v,t)=\,&\frac{1}{\pi}\int_{\mu_b}^\infty d\mu \frac{ \del_t{\text{Abs}}_s \T^+_{\tau_1 \tau_2 }(\mu,t) }{ (\mu - 2m^2+t/2-v)^{N_S+1} } + \frac{\del_t{\text{Abs}}_u \T^+_{\tau_1 \tau_2 }(4m^2-t-\mu,t) }{ ( \mu -2m^2+t/2+v)^{N_S+1} } \\ -\, \frac{N_S+1}{2}\,&\frac{1}{\pi}\int_{\mu_b}^\infty d\mu \frac{ {\text{Abs}}_s \T^+_{\tau_1 \tau_2 }(\mu,t) }{ (\mu - 2m^2+t/2-v)^{N_S+2} } + \frac{  {\text{Abs}}_u \T^+_{\tau_1 \tau_2 }(4m^2-t-\mu,t) }{ ( \mu -2m^2+t/2+v)^{N_S+2} }\, ,
\end{aligned}
\end{equation}
we know by \eqref{eq.posabs} that the first integral of the RHS is positive for $0\leq t<m^2$. The second line looks like $f_{\tau_1\tau_2}(v,t)$, but with an additional power of the denominator. Noting the integral inequality for any positive definite function $\rho(\mu)$
\begin{equation}
    \int_{\mu_b}^\infty d\mu \frac{ \rho(\mu) }{ (\mu - 2m^2+t/2\pm v)^{N_S+2} } < \frac{1}{\M^2} \int_{\mu_b}^\infty d\mu \frac{ \rho(\mu) }{ (\mu - 2m^2+t/2\pm v)^{N_S+1} }\, ,
\end{equation}
for
\begin{equation}
    \M^2 = {\rm Min}_{ \mu \geq \mu_b}(\mu - 2m^2+t/2\pm v)= \mu_b - 2m^2+t/2\pm v\, ,
\end{equation}
we conclude that
\begin{equation}
   \del_t f_{\tau_1\tau_2}(v,t)+\frac{N_S+1}{2\M^2}f_{\tau_1\tau_2}(v,t)>0 \, .
\end{equation}
Similar arguments can be used to obtain bounds on higher derivatives in $t$.

Our bounds are valid, in general, for the ranges of parameters
\begin{equation}
    |v|<2m^2+t/2, \; 0\leq t<m^2, \; \M^2=2m^2+t/2\pm v\, ,
\end{equation}
and at tree level, with $\Lambda$ the EFT cutoff, for the ranges
\begin{equation}
    |v|<\Lambda^2, \; 0\leq t<m^2, \; \M^2 = \Lambda^2 - 2m^2+t/2\pm v\approx\Lambda^2\, .
\end{equation}

\section{Linear Vector Bounds}\label{app.linbounds}

In this part, we express a set of 10 bounds out of the analysis of section~\ref{sec.bounds}.
First, we recall the definition of the main quantities we use here,
\begin{equation}
    f_{\tau_1\tau_2}(v,t)=\frac{1}{N_S!}\frac{\del^{N_S}}{\del s^{N_S}}\tilde\T^+_{\tau_1\tau_2}(s=v+2m^2-t/2,t)\, ,
\end{equation}
\begin{equation}
    f_{\alpha\beta}(v,t) = \sum_{\taus}\alpha_{\tau_1}\beta_{\tau_2}\alpha^*_{\tau_3}\beta^*_{\tau_4}f_\taus(v,t)\, ,
\end{equation}
where $\T^+_\taus$ is the combination of spinning amplitudes in transversity basis defined in \eqref{eq.tplus}. This work focuses on spin-1 theories such that
\begin{equation}
    N_S = 4S+2= 6.
\end{equation}
The bounds can be summarized as follows\footnote{Some higher $t$-derivative bounds could also be considered, but are not relevant here.}.
\begin{align}
    &\del_v^{2N} f_{\tau_1\tau_2}(v,t)>0 \quad \forall\quad N\geq 0\, , \label{eq.sumbound1} \\
    &\del_t f_{\tau_1\tau_2}(v,t)+\frac{N_S+1}{2\M^2}f_{\tau_1\tau_2}(v,t)>0\, , \label{eq.sumbound2} \\
    &\del_v^{2N}f_{\alpha\beta}(0,0)>0 \quad \forall \quad N\geq 0\, , \label{eq.sumbound3}
\end{align}
where, at tree level,
\begin{equation}
    |v|<\Lambda^2, \; 0\leq t<m^2, \; \M^2=\Lambda^2.
\end{equation}

Furthermore, the quantities we are going to find for the GP and PN models are linear in $t$ and $v$. They may be written in the schematic form
\begin{equation}\label{eq.fdef}
\begin{aligned}
    f_{\tau_1\tau_2}(v,t)&=f_{\tau_1\tau_2}(0,0)+ \del_t f_{\tau_1\tau_2}\cdot t\\ &\equiv \frac{1}{\Lambda_2^4}\left[ \mu_{\tau_1\tau_2}+\lambda_{\tau_1\tau_2}\frac{t}{m^2}\right],
\end{aligned}
\end{equation}
and
\begin{equation}
    f_{\alpha\beta}(v,t)=f_{\alpha\beta}(0,0)+\del_t f_{\alpha\beta}\cdot t +\del_v f_{\alpha\beta}\cdot v\, ,
\end{equation}
where $f(0,0)$, $\del_t f$, and $\del_v f$ are linear combinations of the EFT parameters with no kinematics dependence left, where we introduce $\mu$ and $\lambda$ for the bounds that follow.
These are statements coming from our explicit results rather than from any theoretical inputs\footnote{The term in $v$ may seem surprising. Indeed, if there is still an $s$ (hence $v$) dependence after taking $N_S$ derivatives, it means that the amplitude was not respecting the $s^2$ growth of the Froissart bound. That is alright, as we are working with an EFT, and only indicates that unitarity may be perturbatively broken. Our amplitudes in the definite transversity polarization basis do not have such a dependence left.}.

Then, the only non-trivial bounds are the $N=0$ $v$-derivative bounds in \eqref{eq.sumbound1} and \eqref{eq.sumbound3}, and the first $t$-derivative bound \eqref{eq.sumbound2}. Moreover, the second term of the latter is suppressed by a factor of $m^2/\Lambda^2$, but is assured to be strictly positive by \eqref{eq.sumbound1}. Therefore, neglecting this term, we obtain a non-strict inequality for $\del_t f_{\tau_1\tau_2}$, later denoted by $\lambda_{\tau_1\tau_2}$. Explicitly, the remaining bounds are
\begin{equation}\label{eq.b1}
      f_{\tau_1\tau_2}(v,t)
         =f_{\tau_1\tau_2}(0,0)+ \del_t f_{\tau_1\tau_2}\cdot t
         >0\, , \quad 0\leq t<m^2,
\end{equation}
and
\begin{equation}\label{eq.b23}
    \begin{aligned}
         &\del_t f_{\tau_1\tau_2}(v,t)\geq 0\, , \\
         &f_{\alpha\beta}(0,0)>0\, .
    \end{aligned}
\end{equation}
Note that \eqref{eq.b1} is included in these two last bounds (as the definite polarizations are just particular cases of indefinite ones, and $t\geq 0$). Hence, we can focus on the bounds in \eqref{eq.b23}, which have no dependence in $t$ or $v$.

There are four independent quantities for the definite elastic polarizations
\begin{equation}
    \begin{aligned}
        &f_{SS}\equiv f_{00}\, ,\\
        &f_{SV}\equiv f_{0+1}= f_{0-1}=f_{+10}= f_{-10}\, ,\\
        &f_{V_+}\equiv f_{+1+1}=f_{-1-1} \, , \\
        &f_{V_-}\equiv f_{+1-1}=f_{-1+1} \, ,
    \end{aligned}
\end{equation}
where we use $S$ and $V$ to denote 0- and 1-transversity modes respectively. This follows the convention used by \cite{Cheung2016}\footnote{Note that they were working in helicity basis, such that $S$ and $V$ respectively denote 0- and 1-helicity modes in this paper.}. In the following, $\mu_{SS}=f_{SS}(0,0)$, $\lambda_{SS}=\del_t f_{SS}$ and similarly for the other polarizations.
Thus, the $t$-derivative bound correspond to four distinct bounds
\begin{equation}
    \lambda_{SS}\geq 0\, , \quad \lambda_{SV}\geq 0\, ,  \quad \lambda_{V_+}\geq 0\, , \quad \lambda_{V_-}\geq 0\, .
\end{equation}

To study the indefinite polarization bounds we introduce the quantities $\alpha_{\pm1,0}$, $\beta_{\pm1,0}$, which designate the projections along the definite polarization vectors in the transversity basis. Also,
\begin{equation}
    \alpha_\pm \equiv \frac{1}{\sqrt{2}}(\alpha_{-1}\pm\alpha_{+1})\, ,\quad \beta_\pm \equiv \frac{1}{\sqrt{2}}(\beta_{-1}\pm\beta_{+1})\, ,
\end{equation}
where we have the normalization condition
\begin{equation}
    |\alpha|^2= |\alpha_-|^2+|\alpha_0|^2+|\alpha_+|^2=1\, ,
\end{equation}
and similarly for $\beta$.
Then, we explicitly find that the indefinite polarization $f_{\alpha\beta}(0,0)$ can be written in the following form for both GP and PN
\begin{equation}\label{eq.fab}
\begin{aligned}
    f_{\alpha\beta}(0,0)=\frac{1}{\Lambda_2^4}\Big[&\tilde\mu_1|\alpha_+|^2|\beta_+|^2 \\
    +&\tilde\mu_2\big[|\alpha_+|^2(1-|\beta_+|^2)+|\beta_+|^2(1-|\alpha_+|^2)\big] \\
    +&\tilde\mu_3\big[|\alpha_-|^2|\beta_0|^2+|\alpha_0|^2|\beta_-|^2\big] \\
    +&\tilde\mu_4\big[|\alpha_-|^2|\beta_-|^2+|\alpha_0|^2|\beta_0|^2\big]\\
    +&2(\tilde\mu_3-\tilde\mu_4)\Re(\alpha_-\alpha_0^*)\Re(\beta_-\beta_0^*)\\
    +&\tilde\mu_5\big[\Re(\alpha_-\alpha_+^*)\Re(\beta_-\beta_+^*)-\Re(\alpha_0\alpha_+^*)\Re(\beta_0\beta_+^*)\big]\Big]>0\, ,
\end{aligned}
\end{equation}
where the $\tilde\mu$'s are combinations of the EFT coefficients\footnote{This expression for $f_{\alpha\beta}$ is equivalent to (4.4) of \cite{de_Rham_2019}, where we redefined their coefficients in the following way $\tilde\mu_1=\mu_1+2\mu_2$, $\tilde\mu_2=\mu_2$, $\tilde\mu_3=\mu_4$, $\tilde\mu_4=\mu_5$ and $\tilde\mu_5=\mu_3$.}. The quantity $f_{\alpha\beta}$ has to be positive for any polarization. In particular, we focus on some specific ones to deduce some bounds. First, picking $|\alpha_+|^2=|\beta_+|^2=1$ implies, by the normalization condition, that only the first line is non-zero, such that \eqref{eq.fab} reduces to $\tilde{\mu_1}>0$. Similarly, picking $|\alpha_+|^2=1$, $|\beta_+|^2=0$ implies that $\tilde{\mu_2}>0$, $|\alpha_-|^2=|\beta_0|^2=1$ (or $|\alpha_0|^2=|\beta_-|^2=1$) implies that $\tilde{\mu_3}>0$, and $|\alpha_-|^2=|\beta_-|^2=1$ (or $|\alpha_0|^2=|\beta_0|^2=1$) implies that $\tilde{\mu_4}>0$.
In summary, we obtain the following bounds
\begin{equation}\label{mubounds}
    \tilde\mu_1>0\, , \quad \tilde\mu_2>0\, , \quad \tilde\mu_3>0\, , \quad \tilde\mu_4>0\, .
\end{equation}
A condition on $\tilde\mu_5$ is less straightforward to obtain. In order to obtain the most stringent bound one could minimize $f_{\alpha\beta}$, e.g. numerically \cite{Cheung2016,Wang2021}.
Whereas for spin-2 theories only a numerical approach is available, for spin-1 theories an analytic minimazation is conceivable. It has been worked out for the gauge-bosons in the Standard Model EFT in  \cite{Zhou2019,Zhou2021}. It would be interesting to see how to apply their analytical procedure in our models.

Here, we want to keep a simple analytical approach and aim to obtain only sufficient conditions on $\tilde\mu_5$\footnote{ A similar philosophy is adopted in \cite{de_Rham_2019}, where they derive a slightly stronger sufficient condition than ours in their Appendix C. We rather use ours, which is given by the bound on definite polarizations, as it has a more straightforward physical meaning.}. We establish these conditions by looking at the bounds for the definite polarizations $f_{\tau_1\tau_2}(0,0)>0$. Indeed, by considering the $\mathcal{O}(\alpha^2_{\tau_1}\beta^2_{\tau_2})$ in $f_{\alpha\beta}$, we see that
\begin{equation}\label{eq.defindef}
\begin{aligned}
    &\mu_{SS}=\tilde\mu_4>0\, , \\
    &\mu_{SV}=\frac{1}{2}[\tilde\mu_2+\tilde\mu_3]>0\, ,  \\
    &\mu_{V_+}=\frac{1}{4}[\tilde\mu_1+2\tilde\mu_2+\tilde\mu_4+\tilde\mu_5]>0\, , \\ &\mu_{V_-}=\frac{1}{4}[\tilde\mu_1+2\tilde\mu_2+\tilde\mu_4-\tilde\mu_5]>0\, .
\end{aligned}
\end{equation}
Then, the bounds we are going to use for $\tilde\mu_5$ are the following
\begin{equation}
    |\tilde\mu_5|< \tilde\mu_1+2\tilde\mu_2+\tilde\mu_4 \quad\Longleftrightarrow\quad \mu_{V\pm}>0\, .
\end{equation}
As already mentioned, stricter bounds could be found, in particular using numerical methods.
Furthermore, the bounds \eqref{mubounds} derived from the indefinite polarizations are respected by the definite bounds (as expected) on $\mu_{SS}$ and $\mu_{SV}$, but also provide additional information that would not be available from the definite polarization analysis alone. Namely, the bound on $\tilde\mu_1$, and the positivity of $\tilde\mu_2$ and $\tilde\mu_3$ not only as a sum, but also separately.

In sum, we have derived a total of 10 bounds:
\begin{equation}\label{eq.sumbounds}
\begin{aligned}
    &\tilde\mu_1>0\, , \quad \tilde\mu_2>0 \, ,\quad \tilde\mu_3>0\, , \\
    &\mu_{SS}>0\, ,\quad \mu_{V_+}>0\, , \quad \mu_{V_-}>0\, , \\  &\lambda_{SS}\geq0\, , \quad \lambda_{SV}\geq0\, , \quad \lambda_{V_+}\geq0\, , \quad \lambda_{V_-}\geq0\, ,
\end{aligned}
\end{equation}
where the $\mu$'s are bounds in the the forward limit, those with $\tilde{\mu}$'s come from indefinite polarizations, and the $\lambda$'s correspond to the $t$-derivative bounds, which are available due to the analysis beyond the forward limit.

\acknowledgments
LH is supported by funding from the European Research Council (ERC) under the European Unions Horizon 2020 research and innovation programme grant agreement No 801781 and by the Swiss National Science Foundation grant 179740.
CdR thanks the Royal Society for support at ICL through a Wolfson Research Merit Award. CdR is supported by the European Union's Horizon 2020 Research Council grant 724659 MassiveCosmo ERC--2016--COG, by a Simons Foundation award ID 555326 under the Simons Foundation's Origins of the Universe initiative, `\textit{Cosmology Beyond Einstein's Theory}' and by a Simons Investigator award 690508. CdR is also supported by STFC grants ST/P000762/1 and ST/T000791/1.

\bibliographystyle{JHEP}
\bibliography{biblio}
\end{document}